\documentclass[AEJ]{AEA}
\usepackage{amssymb}
\usepackage[pdftex]{graphicx}
\usepackage{fancyhdr}
\usepackage{amsmath}
\usepackage{setspace}
\usepackage{natbib}

\newtheorem{ass}{Assumption}
\newtheorem{prop}{Proposition}
\newtheorem{proof*}{Proof}
\newtheorem{lemma}{Lemma}

\draftSpacing{1.5}

\begin{document}

\title{Optimistic versus Pessimistic\\
        ---Optimal Judgemental Bias with Reference Point}
\shortTitle{Optimistic versus Pessimistic}
\author{Si Chen\thanks{Correspondence address: Department of Economics, University of Oxford, Manor Road Building, Manor Road, Oxford OX1 3UQ, UK}}%,This thesis was written in partial fulfilment of the requirements for the degree of M.Phil. in Economics at University of Oxford. I am deeply indebted to my supervisor, Professor Vincent Crawford for his continuous patience, guidance and support while developing this project. I would also like to thank Professor Sujoy Mukerji, Dr. Margaret Meyer, Dr. Han Ozsoylev and Dr. Mungo Wilson, for their useful comments. Further thanks goes to my schoolfellows:Yixiong Wang, Wen Xu, Kwon Hee Youn.}}
\date{\today}
\pubMonth{July}
\pubYear{2013}
\pubVolume{Vol}
\pubIssue{Issue}
\JEL{}
\Keywords{}

\begin{abstract}
This paper develops a model of reference-dependent assessment of subjective beliefs in which loss-averse people optimally choose the expectation as the reference point to balance the current felicity from the optimistic anticipation and the future disappointment from the realization.The choice of over-optimism or over-pessimism depends on the real chance of success and optimistic decision makers prefer receiving early information. In the portfolio choice problem, pessimistic investors tend to trade conservatively, however, they might trade aggressively if they are sophisticated enough to recognise the biases since low expectation can reduce their fear of loss.
\end{abstract}

\maketitle
Neoclassical economics uses a natural simplification of human behaviour as governed by limitless cognitive ability applied to a handful of perceptible goals and untangled by emotions(\cite{clark98}). One fundamental proposition from this presumption is rational expectation(\cite{muth61}) simply declaring that agents' predictions of the future value of economically relevant variables are not systematically wrong in that all errors are random. Equivalently, this is to say that agents' expectation is consistent with the true statistical expectation. Nevertheless, there are plentiful evidences of deviations from rational expectations, with the most prominent one showing that individuals err in their probability assessments and not in random, but in systematic directions. In particular, when the task is very difficult with rare positive events, people often exhibit over-pessimism and overestimate the probabilities of bad outcomes(\cite{kruger99, wks03, KrugB04}). On the other hand, when the task is easy and the probability of success is reasonably high, people tend to exhibit over-optimism and overestimate the probabilities of good outcomes(\cite{FSL77, Sven1981, Hoff04})\footnote{See \cite{Moore08} for a review.}. In short, subjective belief of a good outcome increases with its real chance but always contains systematic biases.

These systematic biases provide strong evidences against the prevailing assumptions that people are either completely rational or completely irrational and erratic in making predictions toward the future(\cite{DeLong90, Friedman05}). Specifically, people are capable of capturing the probabilities of success and failure to a reasonable extent of accuracy, like rational individuals\footnote{In situations with ambiguity, this statement is still valid(see \cite{Camerer92} for a review). Although agents facing ambiguity are unable to capture the true distribution, they can apprehend a family of such distributions and a probability distribution over the family. Their best guess works as a good proxy of the objective distribution.}; but when it comes to forming subjective beliefs, they will deceive themselves and choose biased beliefs as if they are irrational since utilities are not only from an accurate prediction. Based on these observations, we build a model in which agents choose the optimal subjective beliefs and optimal expectations as in \cite{Brun05, Brun07}\footnote{\cite{Spiegler08} discussed the problems in revealed-preference approach and information acquisition raised by \lq\lq utility-from-belief\rq\rq\ models.}. Different from their models, individuals are reference dependent and loss averse. The expectation formed by subjective beliefs serves naturally as the reference point(\cite{KR06}). Hence, higher anticipation not only produces current felicity as in BP, but also results in potential greater future loss. Individuals manipulate their beliefs to trade off the benefit of an optimistic future with the cost of a painful loss. In particular, my model contains three main elements: 

First, utilities consist of two parts: anticipatory utility and future gain-loss utility. Previous studies suggest that a forward-looking decision maker cares about both utilities(\cite{Loew87, Caplin01, Tirole04,Bernheim05})\footnote{\cite{Caplin04, Spiegler06} pointed out the problems in reveal preference approach brought by forward-looking utility.}. Decision makers weigh anticipatory utility against future utility depending on the specific circumstances such as the time to payoff date and how significant they value the lottery.

% An optimistic expectation increases the anticipatory utility, however, the future utility is determined by comparing with expectation, therefore higher expectation as a higher reference point decreases the reference dependent future utilities. This trade-off lead to two options for an decision maker: she can either be over-optimistic, i.e., setting a high expectation and exposing herself to possible painful loss in the future; or pessimistic, i.e. setting a low expectation and giving herself better chance of enjoying a gain in the future, depending on the total utilities associated with two choices. 

Second, utility is reference dependent. We assume the future utility is reference dependent(\cite{KahnTver79, KR06, KR09}) and the anticipation serves naturally as the reference point(\cite{KR06}). Anticipatory and future utilities are therefore linked together by the reference point which is constituted by an agent's subjective beliefs. For simplicity, anticipatory utility is assumed to be reference independent and it takes the value of ordinary expectation.

%\footnote{The anticipatory utility can be either reference dependent or independent depending on the real economic settings we are dealing with. Koszegi and Rabin(2009)took the reference-dependent assumption while Bernheim and Thomadsen (2003) simply assumed anticipation is reference independent. Intuitively, anticipatory utility is reference dependent if agents have recent or abundant experiences that significantly affect their current feelings, e.g. anticipation in repeated experiments and different felicities from the same anticipation for an amateur and a professional chess player.} 
%In the following analysis,Otherwise, 

Third, individuals are loss averse. Previous studies without loss-aversion usually lead to optimistic judgemental biases(\cite{Brun05, Brun07, Bernheim05}). In our model, over-pessimism is also possible, because loss due to optimism can be more heavily felt than the felicity from a good anticipation.

All these three factors together give our decision makers inner conflicts in making predictions. They always enjoy being ambitious, while hate to be disappointed; enjoy feeling lucky, while hate to be hopeless. Optimal beliefs (and associated optimistic and pessimistic attitudes) are then chosen by weighting current anticipation and future utility, depending on the probabilistic properties of the lotteries, the future discount rate and the intensity of loss aversion.

%To set beliefs towards the future, agents take all these current and future feelings into consideration. Expectation is then chosen by weighing the gains versus losses, leading to "rational" biases of either over-optimistic or over-pessimistic, depending on the real probabilistic properties of the lotteries, the discount of future utilities and the intensity of loss aversion.

Psychological theories and evidence support this intuition. Contemporary psychology acknowledges that the human behaviour is influenced simultaneously by conscious(control System 1) and nonconscious(automatic, System 1) processes(\cite{Chai99}). Neuropsychology using fMRI technique has provided evidence that the ventral medial prefonrtal cortex compete with the prefrontal cortex for the control of the response to the problems given in the belief-bias test(\cite{Goel03}). These studies indicate that individuals are more capable than previously assumed in capturing the real probabilistic pattern of stochastic events, but can subconsciously deceive themselves into biased beliefs. To quote \cite{Kahneman11}, the process through which people choose subjective beliefs is the result of an uneasy interaction between two systems: \lq\lq System 2 articulates judgements and makes choices, but it often endorses or rationalizes ideas and feelings that were generated by System 1.\rq\rq . Even though the real biological process in distorting beliefs is not fully deciphered, this view gives a reasonable interpretation that human mind is operated as a dual and conflicting system---pursuing mental pleasures like idealists while staying close to reality as physicalists. 

Moreover, previous experimental studies pointed out that subjective beliefs can change from optimism to pessimism without any additional information(\cite{shepperd96, shepperd98, shepperd06, Mayraz11b}) For example, \cite{shepperd96} conducted an experiment in which students estimate their exam scores a month before the exam, then again several times after completing the exam yet prior to receiving feedbacks. As the date of feedback drawing near, students turn their optimistic forecasts into pessimistic forecasts. The switch in attitudes implies that people can manipulate beliefs for their interests: when the threaten of loss becomes more relevant, pessimism is preferred to optimism since it placates the anxiousness of loss. In \cite{Mayraz11b}, price predictions of financial assets based on the same historical charts are significantly higher for subjects gaining from high prices than those gaining from low prices. Asymmetry in predictions is attributed to the different choices of anticipations as reference points.  

%Moreover, Michele Piccione and Ariel Rubinstein (1997) stated that individuals understand that they have imperfect recall, and in B. Douglas Bernheim and Raphael Thomadsen (2005), individuals additionally can influence the memory process to increase anticipatory utility. \footnote{Futhermore, in Yariv(2002), Matthew Rabin and Joel L. Schrag(1999), beliefs are distorted to be consistent with past choices.}\\

This paper proceeds as follows. After presenting the relevant literatures in Section I, Section II presents the agent's formal problem and model predictions on the choice of optimal beliefs, starting with the general multiple-state model followed by a two-state example. Subjective beliefs bias up and down depending on whether the real chance of success exceeds a cut-off probability uniquely determined by intensity of loss aversion and future discount rate. We further discuss a short application on information timing preference following the work in \cite{KR09}(KR2009). It is shown that as the confidence level decreases, biased agents abandon their preference to early information in favour of staying uninformed while rational agents are indifferent. Section III explores the basic premise of the model---risk attitudes over lotteries. We divide the agents into two types---\lq\lq naive\rq\rq\ and \lq\lq sophisticated\rq\rq\ , according to their different cognitive processes: sophisticated agents recognise their biases in making decisions while naive ones do not. Therefore, naive agents show preference to skewness and spreading in distribution while sophisticated agents behave similarly to rational ones except overweighing low-ranking outcomes. Section IV applies the model to the portfolio choice problem. Following previous categorisation of agents, we conclude that naive optimistic agents trade aggressively while pessimistic ones trade conservatively. Sophisticated agents can bear more(less) risks than rational ones when low-ranking returns\footnote{Low-ranking returns refer to the bad returns of a risky asset.} are good(bad) relative to risk-free return, because loss aversion dictates a higher weight on bad returns. Moreover, sophisticated pessimistic investors can bear excess risks than optimistic ones because low anticipation reduces their fears for loss and makes them numb. Furthermore, we explore the pricing implication in a market with identical investors and short-sale constraint. Price of the risky asset in a naive market decreases as the market becomes pessimistic but exhibits a U-shape in a sophisticated market implying that a market with the moderate confidence level has the highest equity premium. Finally, Section V discusses the model's scope and limitations. 
\section{Relevant Literature}
Related literatures are divided into three groups: 1.optimal beliefs 2.reference-dependent utilities 3.biased beliefs in asset pricing. 
\subsection{Optimal beliefs}
Literatures on distorted optimal beliefs are built on the assumption that people choose subjective biased beliefs departing from the real probabilities. Previous studies can be divided into three branches: one focuses on anticipatory utilities of forward-looking decision makers\footnote{\cite{Akerlof82} proposed a model in which workers in hazardous professions choose their subjective beliefs of an accident to balance their anticipatory feelings of danger and money spent on safety equipments.}; one focuses on cognitive dissonance in which people hold inconsistent beliefs to comfort their past experiences(\cite{Rabin94, Larry06})\footnote{For example, \cite{Larry06} built a axiomatic model in which agents adjust their beliefs after taking an action so as to be more optimistic about the possible consequences.}; the third branch focuses on biased beliefs arising from self-signalling with imperfect memory(\cite{Tirole02,Tirole04,Bernheim05})\footnote{For example, \cite{Bernheim05} developed multiself-consistency game where decision makers have both anticipatory utility and imperfect memory. In order to benefit from anticipation, agents with imperfect recall prefer staying uninformed and exhibit over-optimism.}. This review focuses on the first branch that is most relevant to the thesis.

\cite{Brun05}(BP2005) built a structural model and gave two applications: choice between a risky and a risk-free asset and a consumption-saving problem with stochastic income. The underlying intuition is straightforward: agents with anticipatory utilities are willing to hold optimistic biased beliefs to achieve higher current felicity. They trade off the current felicity from a higher expectation with the cost of making a sub-optimal decision and thus worse-realized outcomes due to the biased beliefs. Agents balance these effects and choose their optimal beliefs to maximize the average utility. They concluded that a small optimistic bias in beliefs typically leads to a first-order gain in anticipatory utility while only induces second-order cost from the poor decision. Further applying this model to the financial market, BP achieved two conclusions: 1. Investors always overestimate the return of their investment, which encourages them to long or short too much of the risky asset compared to what would maximize their objective expected utility\footnote{Specifically, they enter into over-possession of the risky asset with average excess return greater than 0, and hold inadequate risky asset with average excess return smaller than 0.}. 2. Investors tend to invest in an asset with high level of positive skewness even if the asset earns a negative average excess return. In the consumption-saving application, BP further concluded that agents are both over-confident and over-optimistic. 

\cite{Spiegler08} criticized BP's model in two perspectives : 1. BP's model fails the rudimentary revealed preference test since the IIA (Irrelevance of Independent Alternatives) is violated. The reason is that subjective beliefs in BP’s model are derived directly from payoffs of the lottery in the choice set\footnote{\cite{Spiegler08} constructed 3 lotteries in states $s_1 ...s_n$ with payoffs in the matrix: \\
$\begin{array}{ccccc}
action/state & s_{1} & s_{2} & ... & s_{n}\\
l_{f} & 0 & 0 & 0 & 0\\
l_{r} & 1 & -k & -k & -k\\
l_{r'} & m & -n & -n & -n
\end{array}$, \\
where $k,m,n>0, m>1$ and $k$ satisfies: $c_{BP}\{l_f,l_r\}=c\{l_f,l_r\}$. Based on BP's model, subjective belief to $s_1$ is $q_1=1$. When the choice set becomes $\{l_f,l_r,l_{r'}\}$, $q_1=1$ leads to $l_{r'}\succ l_{r}$ and $c_{BP}\{l_f,l_r, l_{r'}\}=l_{r'}$. However, with $n$ big enough, $l_{f}\succ l_{r'}$,leading to a violation of IIA.}. 2. BP's model cannot capture the preference for biased information sources. Intuitively, by assuming people have desires to attain self-serving beliefs, BP's model should also provide explanations of people's preference for information sources which can distort the beliefs indirectly. However, Spiegler proved that in BP, the decision maker is never averse to information since the support of subjective beliefs does not update to signals containing uncertainty. The rationale for this criticism lies in BP's assumption that the action, rather than beliefs is the most fundamental choice variable. Subjective beliefs are inertial to new information as long as the punishment from the sub-optimal action stays the same. With the assistant of loss aversion, decision makers in our model are able to choose subjective beliefs directly and they exhibit different tastes for information depending on their optimistic and pessimistic attitudes. Experimental evidence also fails to justify BP’s conclusions. \cite{Mayraz11b} gives a further critique on BP2005: BP's model assumes that the biased beliefs are costly for a decision maker in material terms due to poor decisions. That means, the magnitude of the bias depends on the incentives for accuracy, and the bias can only be substantial when incentives for accuracy are weak. However, this assumption is rejected by experimental data in \cite{Mayraz11b}: it is observed that the biases are independent of material costs from poor decisions\footnote{Subjects observed the historical price chart of a financial asset, and received both an accuracy bonus for predicting the price at some future point, and an unconditional award that was either increasing or decreasing in this price. The statistical test indicates that the magnitude of the bias is independent of the amount paid for accurate predictions.}. An extension to Mayraz's intuition is that in situations where the decision maker has no control over final outcomes, there is no punishment for sub-optimal actions and optimal-belief holders in BP will bias up as much as they wish. Finally, Mayraz developed an axiomatic model in which choice of beliefs is payoff dependent: agents believe what they want to be true (\cite{Mayraz11a}). Agents derive their optimistic and pessimistic attitudes based on an attitude parameter exogenous to the model.

Our model mainly takes the framework of BP with some modifications. Like BP, decision makers in our model are forward-looking and the introduction of anticipatory utility to the ordinary economic model brings the tendency of optimism. Unlike BP, decision makers are loss averse and this behavioural element provides first-order counter force to anticipation, making pessimism preferable to optimism when loss aversion is strong enough. In situations where agents do not discount future, pessimism is always more beneficial since the intense hurt from loss by setting a high anticipation always exceeds the happiness from gain brought by that anticipation. 

\subsection{Reference-dependent utilities}
The prospect theory first proposed by Daniel Kahneman and Amos Tversky in their 1979 paper pointed out that: the evaluation of outcomes is compared to a reference point; the aversion to loss is significantly intensive than the felicity from gain with a diminishing sensitivity to changes in an outcome as it moves farther away from the reference point; subjective probability of a prospect is non-linear in the true probability---in particular, people overweigh small probabilities and under-weigh high probabilities. \cite{KR06} built a model with a separation of the reference-independent \lq\lq consumption utility\rq\rq\ and the \lq\lq gain-loss utility\rq\rq\  based on the essential intuition of \cite{KahnTver79}. The reference point in their model is simply people's rational expectation determined by the \lq\lq personal equilibrium\rq\rq\ ---an equilibrium in which expectation is consistent with the optimal choice under that expectation. With application to the consumer behaviour, they observed that the price a person wishes to pay for a commodity increases with the expected price conditional on purchase and the expected probability of purchase. Another application lies in the within-day labour-supply decision. A worker is likely to continue working only if they receive income less than their expectation. Based on their 2006 model, \cite{KR09}(KR2009) developed a rational dynamic model in which people are loss averse over changes in rational expectation about present and future consumption. They concluded that when agents are more sensitive to news about upcoming consumption than to news about distant consumption, then 1. agents prefer receiving early information rather than later; 2. agents boost consumption immediately but delays cuts; 3. Agents feel piecemeal information undesirable due to the diminishing marginal utility of loss. Our model employs KR's setting of utility function: utility is composed of separable reference-independent \lq\lq consumption utility\rq\rq\ and reference-dependent \lq\lq gain-loss utility\rq\rq . Different from their model, however, agents are no longer rational and can optimally manipulate their beliefs. It should be noted that \cite{Spiegler06} commented that \lq\lq the model fails to account for a variety of realistic prior-dependent attitudes to information, which intuitively seems to originate from anticipatory feelings\rq\rq \footnote{For instance, a patient who wants to have full knowledge of her medical condition when she is quite sure that she is in good health, yet does not want to know the whole truth when she is not so sure(\cite{Spiegler06}).}. Finally, \cite{Macera11} explored the time-path of subjective probability assessment. She built a two-state model in which an agent experiences gains and losses from the changes in anticipation and waits T periods for the realisation of the outcome. In each period, the agent makes assessment of her likelihood of success to maximize the intertemporal utility. One major conclusion is that the optimism decreases as the payoff date gets close because the threat of disappointment becomes significant. With application to the design of bonuses, she found that the decreasing path of optimism leads to a strong preference to bonuses. Thus optimal bonuses granted with a periodicity strengthen the motivation while restrain the payoffs in a reasonable size. Different from \cite{Macera11}, our model explores the choice of subjective beliefs in the multi-state setting where ex post dissonance is excluded from discussion.
 
\subsection{Biased beliefs in asset pricing}
\cite{Huang08} studied the asset pricing implications of the cumulative prospect theory in \cite{KahnTver92}, focusing on the probability weighting component. They proved that in a one-period equilibrium setting with normally distributed security payoffs and homogeneous investors evaluating risk based on cumulative prospect theory, the CAPM still holds. Moreover, they introduced an additional small, independent and positively skewed security into the economy and derived an equilibrium for homogeneous cumulative prospect theory investors. As the investors overweigh the tails of the portfolio and skewed security's return, they concluded that the skewed security can be overpriced and can earn a negative average excess return. A further extension to BP2005 is made in \cite{Brun07} in which they built a general equilibrium model with complete markets. They showed that when investors hold optimal beliefs like in BP2005, portfolio choice and security prices match six observed patterns: 1. investors are not perfectly diversified due to biased beliefs; 2. the cost of biased beliefs puts limits on biases and make the utility cost not explosive; 3. Investors only over-invest in one Arrow-Debreu security and smooth the consumptions across all the other states because of the complementarity between believing a state more likely and purchasing more of the asset that pays off in that state; 4. Identical investors can have heterogeneous optimal portfolio choices since different households have different states to be optimistic about; 5. investors tend to over-invest in the most skewed asset because the low-price and low-probability states are the cheapest states to buy consumption in. Thus over-optimism about these states distorts consumption the least in the rest of the states; 6. more skewed assets provide lower returns because of the higher demand for them. 

The structure of subjective beliefs in our model is compatible to the cumulative prospect theory which serves as the foundation of \cite{Huang08}. Further more, the preference to pessimism may lead to different conclusions to \cite{Brun07} in Arrow-Debreu asset pricing model. Though this thesis mainly focus on the fundamental discussions of people's risk attitudes, light can be shed on asset pricing in markets with optimistic or pessimistic attitudes.

%, light can be shed on risky and skewed asset pricing in optimistic and pessimistic markets.   

\section{The Model}
\subsection{The Utility Function}

Consider a multi-state model with two periods and one agent. There is a lottery with contingent payoffs and the associated distribution that are known by the agent. We assume the agent has no control over the true distribution or the final outcomes but she can deceive herself by manipulating her subjective beliefs. The first period corresponds to the time when the agent builds her beliefs and anticipation and the second period is the payoff realization period. The agent at t=1 has an imminent utility from her anticipation of the future payoffs as well as a prospective gain-loss utility due to the difference between the payoffs actually realised at t=2 and her anticipation at t=1. The agent chooses her optimal subjective beliefs by considering both the anticipatory utility and the future prospective gain-loss utility. We further assume that the subjective beliefs remain constant over two periods, meaning that regression is not allowed once beliefs are formed. The model is formally described as following.

$\mathbf{\mathcal{S}}=\{1,...,S\}$, $S\geq2$, is the set of states of nature and  $Z_{s}={Z_{1},...,Z_{S}}$ are the corresponding material payoffs where $0\leq Z_{1}\leq Z_{2}\leq...\leq Z_{S}$. The objective probability associated with the state $s$ is $p_s$, for $s=1,...,S$. Following the basic assumptions made in \cite{KR09}, we assume that an agent's utility is separable and consists of a \lq\lq consumption utility\rq\rq\ and a \lq\lq universal gain-loss utility\rq\rq\ \footnote{In \cite{KR09}, the decision maker's period-t instantaneous utility $u_t$ depends on the consumption in period t, and the changes in period t to beliefs about contemporaneous and future consumption:\\
$$
u_{t}=m(c_{t})+\sum_{\tau=t}^{T}\gamma_{t,\tau}N(F_{t,\tau}|F_{t-1,\tau}).
$$
$m(c_t)$ represents the \lq\lq consumption utility\rq\rq\ and can be thought of as the classical reference-independent utility. $N(F_{t,\tau}|F_{t-1,\tau})=\int_{0}^{1}\mu(m(c_{F_{t,\tau}}(p))-m(c_{F_{t-1,\tau}}(p)))dp$ represents the \lq\lq gain-loss utility\rq\rq\ and is derived from the changes in beliefs over future outcomes between periods  where $c_{ F }(p)$ is the consumption level at percentile $p$ and $\mu(\cdot)$ the universal gain-loss utility function. In our model, we simplify their model into two periods and assume there is no contemporaneous gain-loss utility at $t=1$. Thus anticipation takes the value of expectation. At t=1, there is a prospective gain-loss utility due to the deviation of subjective beliefs from objective ones. We assume $\gamma=1$ since the effect of $\gamma$ can be included in the effect of $\eta$ under our linear setting of gain-loss utility function.}: consumption utility takes the form of $u(Z_s)$, with $u'>0,u''\leq0,u(0)=0,$ and the \lq\lq gain-loss utility\rq\rq\ takes the form of $\mu(x)=x$ for $x\geq0$ and $\mu(x)=\lambda x\,(\lambda>1$) for $x<0$.\footnote{In Appendix B, we discuss the case with a more general assumption $\mu'(-x)=\lambda(x)\mu'(x)$, where $\lambda(x)>1$, $x>0$, and $\underset{x\rightarrow0}{lim}\lambda(x)=1$, $\mu''(x)\leq0$. The conclusions are not different from the linear assumption case.} For simplicity, $u(Z_s)$ is denoted by $u_s$ and we have $u_{1}\leq u_{2}\leq...\leq u_{S}$ by assumption. The anticipatory utility formed by the agent's subjective beliefs about the consumption utility realised in the second period is $\underset{s\in\mathbf{\mathcal{S}}}{\sum}q_{s}u_s,$ where $q_s$ is the subjective beliefs assigned to state $s$ for $s=1,...,S$. The prospective gain-loss utility comes from the difference between the subjective beliefs and the objective ones is $\underset{s\in\mathbf{\mathcal{S}}}{\sum}p_{s}\mu(u_s-\underset{s\in\mathbf{\mathcal{S}}}{\sum}q_{s}u_s).$
The agent at t=1 chooses her optimal subjective beliefs to solve the following maximization problem
\begin{equation}\label{OB}
\underset{\{q_{s}\}_{s\in\mathcal{S}}}{Max}U=\underset{s\in\mathbf{\mathcal{S}}}{\sum}q_{s}u_s+\eta\{\underset{s\in\mathbf{\mathcal{S}}}{\sum}p_{s}\mu(u_s-\underset{s\in\mathbf{\mathcal{S}}}{\sum}q_{s}u_s)\}
\end{equation}
\begin{equation*}
s.t. \quad\underset{s\in\mathbf{\mathcal{S}}}{\sum}q_{s}=1 
\end{equation*}
where $0<\eta\leq1$ is the weight on \lq\lq gain-loss utility\rq\rq\ with the weight on \lq\lq consumption utility\rq\rq\ normalized to 1. The upper bound of $\eta$ is set to 1 to meet the revealed preference requirement.\footnote{Specifically, this assumption is required to meet the revealed preference between these two lotteries: $(0,0)$ and $(1,0)$ with probabilities $(p,1-p)$. Obviously, we have $(1,0)\succeq(0,0)$ for any value of $p$. Therefore, $p+\eta(1-p)-\eta\lambda p\geq 0$. Then we have $\eta\leq\frac{-p}{1-p-\lambda p}$. For $p=\frac{1}{\lambda}$, we can derive that $\eta\leq 1$. }. The weight on gain-loss utility can also be viewed as the discount rate of future utility. \\

\subsection{Optimal Beliefs}
This section presents the fundamental properties of optimal beliefs in \ref{OB}. All the proofs are given in Appendix A. 

\begin{prop}{(Biased Beliefs are Preferred to Rational Unbiased Beliefs):}\label{bsbetter}\\
If $\lambda\geq\dfrac{1}{\eta}$, then for $s=1,...,S, S\geq2$, there exists
at least one $q_{s}\neq p_{s}$, s.t. $U_{BS}\geq U_{RE}$, where
$U_{BS}$ and $U_{RE}$ are the utilities under biased and rational beliefs respectively.\\
\end{prop}

Proposition \ref{bsbetter} says that an agent who cares about both anticipatory utility and prospective gain-loss utility never holds rational beliefs, even though being either over-optimistic or over-pessimistic is costly. The model rationalizes the existence of biased beliefs when agents have no control power over the realisation of a gamble.

\begin{prop}{(Beliefs Tradeoff among Different States):}\label{tradeoff}\\
An agent prefers moving a small probability $\varepsilon$ from a bad(good) state to a bad(good) state if the objective probability of getting an outcome better than her expectation is high(low): \\
$\forall k,l\in\mathbf{\mathcal{S}}$ with $k>l$, if $P_{+}>(<)P^{*}$, then we have, \\
$$U(q_{k}+\varepsilon,q_{l}-\varepsilon,q^{-})>U(q_{k},q_{l},q^{-}),$$
where $\varepsilon>(<)0$
is a small number; $P_{+}=\underset{A}{\sum}p_{s},\: A=\{s\in\mathbf{\mathcal{S}}:\; u_s-\underset{s\in\mathbf{\mathcal{S}}}{\sum}q_{s}u_s\geq0\}$;
$P^{*}=\dfrac{\eta\lambda-1}{\eta(\lambda-1)}$. \\
\end{prop}
\begin{prop}{(Over-optimistic versus Over-pessimistic)}\label{op}\\
Optimal beliefs defined by problem \ref{OB} feature the following properties:\\
(i) The probabilities of high(low) rank outcomes are over-estimated(under-estimated) if the objective probability of getting an outcome better than the objective expectation is high(low).\\ 
That is, on average, an agent is over-optimistic(over-pessimistic):\\
$$\underset{s\in\mathbf{\mathcal{S}}}{\sum}q_{s}u_{s}>(<)\underset{s\in\mathbf{\mathcal{S}}}{\sum}p_{s}u_{s},$$
if $P_{+}^{0}>(<)P^{*},$ where $P=\underset{A^{0}}{\sum}p_{s},\: A^{0}=\{s\in\mathbf{\mathcal{S}}:\; u_s-\underset{s\in\mathbf{\mathcal{S}}}{\sum}p_{s}u_s\geq0\}$.\\
\\
(ii) An agent is less likely to be over-optimistic if she is more loss averse and if she cares more about future utility: $P^{*}$is increasing in $\lambda$ and $\eta$.
 \\
(iii)The optimal set of subjective beliefs $\{q_{s}\}$ satisfies $P_{+}=P^{*}$ and is not unique. \\
\end{prop}

Proposition \ref{tradeoff} and Proposition \ref{op} are related to each other. Proposition \ref{tradeoff} describes the dynamic process of adjusting subjective probabilities among different states, while Proposition \ref{op} illustrates the properties of subjective beliefs and expectations in the stable state. 

A shared factor in the two propositions is the cut-off probability $P^*$ which is uniquely determined by an agent's preference parameters: the intensity of loss-aversion $\lambda$ and the weight on gain-loss utility $\eta$. For any lottery, an agent determines whether she will further increase or decrease her subjective expectation by comparing the total chance of getting a gain with $P^*$. The cut-off rule implies that despite distortions in beliefs, there is a reasonable correspondence between true probabilities and optimal beliefs: the higher probabilities of good outcomes give stronger reason to be optimistic. 
%Consistent with empirical evidence, people tend to  be over-optimistic and overestimate the chance of good outcomes when they are very likely to happen.
Intuitively, people tend to be over-pessimistic and set a low anticipation to reduce the potential painful feeling of loss when the bad outcomes are very likely to happen. On the contrary, when the chance for good outcomes is high, people tend to be over-optimistic and overestimate the chance of good outcomes, because the threaten of loss is relatively weak and high anticipation is more beneficial. Our intuition is consistent with empirical evidence in previous studies described in Chapter 1. 

%In Proposition \ref{tradeoff}, when the probability of gain is higher than $P^*$, then further up-bias is desirable as the increase in anticipatory utility from a higher expectation exceeds the decrease in gain-loss utility by turning a state from gain to loss under the new higher reference point. The marginal decrease in gain-loss utility is determined by the preference parameters. Similar analysis are applicable for down-bias when the probability of gain is smaller than $P^*$. The up and down distortions happen by moving probabilities between any two different states as described by proposition\ref{tradeoff}. 

In stable state, beliefs and expectations are adjusted to a level to make the total probability from \lq\lq gain\rq\rq\ states equal to the cut-off $P^*$. Whether the subjective expectation is above or below the rational expectation is determined by the real probability distribution of the lottery. In particular, the subjective expectation is higher(lower) than the rational level if the total probability of gains under rational expectation is smaller(greater) than $P^*$. 

The second fold of proposition \ref{op} is also very intuitive: a person who is more loss averse and cares more about future utility tends to have low confidence. This intuition is directly reflected in the cut-off $P^*$: higher $P^*$ leaves fewer room for optimism and those agents with high $P^*$ turn into optimistic only if the lottery promises even higher chance of success. This result can be related to the empirical facts \cite{shepperd96, shepperd98} that the level of optimism decreases as the realisation date comes closer, in which case, $P^*$ gradually increases to 1 as the discount factor $\eta$ approaches 1.

Finally, the optimal sets of beliefs are not unique. The cut-off rule only determines whether an agent is over-optimistic or over-pessimistic on average. Beliefs assigned to distribution tails can be opposite to the average trend, that is an optimistic agent can underestimate the chances of extremely good outcomes. Therefore, the structure of subjective beliefs is compatible with the Cumulative Prospect Theory proposed by \cite{KahnTver92}. 

\subsection{A Short Application: Information Timing preference}

The model above describes the process of self-deception---people directly choose their confidence level. Both casual observations and experimental evidence suggest that self-confidence has a lot to do with the information seeking behaviour. Following the work of KR2009 in which they studied rational agents' timing preference of information, this section extends the model to the preference of agents with optimistic and pessimistic biases. 

We follow the assumptions made above on utility functions. Now, the agent may receive an early signal $i\in I=\{1,2,...,S\}$  at t=1 about her future payoff and the signal is always correct, meaning that the payoff realised at t=2 will be no different from what the agent has learned at t=1. Or, an agent can refuse to observe the early signal and wait until payoff is realised at t=2. \\
The agent's expected total utility from observing an early signal is
\begin{equation}\label{Earlysignalgen}
\underset{s\in\mathbf{\mathcal{S}}}{\sum}q_{s}u_{s}+\eta\cdot\underset{s\in\mathbf{\mathcal{S}}}{\sum}q_{s}\mu(u_{s}-\underset{s\in\mathbf{\mathcal{S}}}{\sum}q_{s}u_{s})
\end{equation}
The first term is the anticipatory utility stated as before while the second term captures the expected prospective gain-loss utility in period 1 for an agent holding biased subjective beliefs $\{q_s\}$. As the signal is correct,  no further gain-loss utility occurs in the second period. 
Compared with the previous situation without early signal, the following propositions are proved to be valid (Details of the proof can be found in Appendix A).

\begin{prop}{(Information Timing Preference)}\label{time}\\
An agent prefers (not) to receive early information about her payoff if
the objective probability of getting an outcome better than the expectation based on objective beliefs is high(low). That is, \\
if $P_{+}^{0}>(<)P^{*}$, then $U_{early}>(<)U_{wait}$, where $P_{+}^{0}=\underset{A^{0}}{\sum}p_{s},\: A^{0}=\{s\in\mathbf{\mathcal{S}}:\; u_s-\underset{s\in\mathbf{\mathcal{S}}}{\sum}p_{s}u_s\geq0\}$
and $P^{*}=\dfrac{\eta\lambda-1}{\eta(\lambda-1)}$. \\
\end{prop}

Proposition \ref{time} says that if $P_{+}^{0}>P^{*}$, an agent strictly prefers to receive the information early; if $P_{+}^{0}=P^{*}$, she is indifferent; if $P_{+}^{0}<P^{*}$, she prefers to stay uninformed. Intuitively, an optimistic agent tends to seek early information because she believes a good signal is more likely to happen. Instead, early information is undesirable for a pessimistic agent since she is unwilling to expose to bad results early ex ante. Pessimistic agents prefer the gain-loss utility coming in the period of realisation as they overestimate the chance of loss in advance. 

Furthermore, our conclusion extends the conclusion of KR2009. In KR2009, they proved that a rational agent weighting equally on \lq\lq prospective gain-loss utilities\rq\rq in both periods is indifferent between the early and later information\footnote{\cite{KR09} built a multi-period model in which $\gamma_{t_1,t_2}$ represents the strength of the concern in period $t_1$ for \lq\lq prospective gain-loss utility\rq\rq\  in period $t_2$. The prospective gain-loss utility stems from changes in beliefs between last period and this period in beliefs regarding future outcomes. Since there are only two periods in our model, $\gamma_{t_1,t_2}$ is degenerated to $\gamma$ since $\gamma_{t_1,t_1}=1$ by their assumption.}. (\ref{Earlysignalgen}) is the respective utility function from observing early signal for an agent with constant weight on \lq\lq prospective gain-loss utility\rq\rq  over periods \footnote{Formally, (\ref{Earlysignalgen}) can be written as \begin{equation*}
\underset{s\in\mathbf{\mathcal{S}}}{\sum}q_{s}u_{s}+\eta\cdot \gamma \underset{s\in\mathbf{\mathcal{S}}}{\sum}q_{s}\mu(u_{s}-\underset{s\in\mathbf{\mathcal{S}}}{\sum}q_{s}u_{s})
\end{equation*}. When $\gamma=1$, it becomes (\ref{Earlysignalgen})}. Their intuition is: a rational agent is unbiased in the probabilities with which an early signal will move beliefs up and down. When the sense of loss for immediate and non-immediate outcomes is equally aversive, the rational agent is indifferent between early and late information. On the contrary, with freedom in choosing subjective beliefs, agents exhibit a preference on information timing depending not only on their discount factor over future, but also on the distribution of the lottery they are playing with. 

\subsection{An Example}

This section works out a simple two-state example, to serve as an antidote to the abstractness of the previous section. In the example, there are two possible outcomes, $x=0,1$ and $u(0)=0, u(1)=1$ with objective probability $p$ and $1-p$ respectively. By holding subjective beliefs $q$ and $1-q$, the anticipatory utility at t=1 is,
$$
U_{A}=E_{q}u(x)=qu(1)+(1-q)u(0)
$$
The prospective gain-loss utility from realised outcomes at t=1 is, 
$$
U_{R}=E_{p}\mu(u(x)-U_A)=p\mu(1-q)+(1-p)\mu(-q)
$$
Total utility under subjective beliefs is,
\begin{equation*}
U_{BS}=\eta p+(1-\eta\lambda)q+(\eta\lambda-\eta)pq;
\end{equation*}
Instead, an unbiased agent has
\begin{equation*}
U_{RE}=\eta p+(1-\eta\lambda)p+(\eta\lambda-\eta)p^{2}
\end{equation*}

It is easy to derive from here that the cut-off $P^*=\dfrac{\eta\lambda-1}{\eta(\lambda-1)}$, which is increasing in $\eta$ and $\lambda$. An agent chooses $q>p$, thus is over-optimistic if and only if $p>P^*$. Otherwise, she chooses $q<p$ and stays in over-pessimistic if $p<P^*$. At $p=P^*$, $U_{RE}=U_{BS}$ for any $q$.

%$P^*$ is increasing in $\eta$. The setting of consumption utility implicitly indicates that the intensity of feeling from gain-loss utility compared to consumption utility is $\eta$. $\eta$ could also be viewed as the weight on gain-loss utility at t=2 when the weight on utility at t=1 is 1 or the discount rate. Intuitively, people who care more about current felicity from high anticipation would certainly be more likely to biased up. 

%$P^{*}$is increasing in $\lambda$. The cost of a high anticipation increases as an agent becomes more loss-averse since pain of loss is more heavily felt. Therefore, highly loss averse agent are less likely be over-optimistic.

Different from the multiple-state case, the discrete two-state setting has $U_{BS}$ maximized at $q=1$ for $p>P^{*}$, and at $q=0$ for $p<P^{*}$ since $p$ rarely equals $P^*$

Now, we consider the agent's timing preference of information. 

The agent may receive an early signal $i=\{0,1\}$ about her future outcome at $t=1$ and the signal is always correct.

%By observing $i=1$, the agent at $t_1$ knows that $x=1$ will happen at $t_2$ and total utility by observing i=1 is,\\
%$$ U_{1}=1+\gamma\eta(1-q), $$

%The explanation is, at $t_1$, by observing i=1, the agent has an anticipatory utility $u(1)$ with certainty and a prospective gain compared to her prior subjective belief q. At $t_2$, as the agent has already updated their reference point to 1, there will be no more gain-loss utilities in this period.

%Instead, if $i=0$, in which case $x=0$ will happen at $t_2$ and the total utility at $t_1$ is,\\
%$$
%U_0=-\gamma \eta \lambda q. 
%$$
For an agent holding optimal biased beliefs, total utility by observing the early signal is,\\
$$
U_{early}=q+\gamma\eta q(1-q)-\gamma\eta\lambda q(1-q), 
$$\\
where $\mathbf{\gamma}$ is the weight on prospective gain-loss utility following the assumption of KR2009. 

At $t=1$, the agent holding biased beliefs believes that with probability $q$, she is going to observe $i=1$---leading to an anticipatory utility $u(1)$ with certainty and a prospective gain compared to her prior q, and, with probability $1-q$, she is going to observe $i=0$---leading to an anticipatory utility $u(0)$ with certainty and a prospective loss. At t=2, as the agent has already updated her reference point to the right level, there will be no more gain-loss utilities in this period.

Instead, if the agent does not observe the signal, her total utility is the same as before: 
$$
U_{wait}=q+\eta p(1-q)-\lambda\eta q(1-p). 
$$
Hence, when $\gamma=1$ as in KR, observing the signal generates strictly more expected utility than not observing it if and only if 
\begin{equation}\label{d}
(q-p)(1-q)>\lambda(p-q)q.
\end{equation}

Since only one side of (\ref{d}) can be greater than 0, we have, \\
$U_{early}>U_{wait}$ if and only if $q>p$, that is $p>P^{*}$. Otherwise, if $q<P^*$, then $U_{early}<U_{wait}$, the agent will prefer staying uninformed. \\

This conclusion is intuitive. People being over-optimistic over payoffs are also over-optimistic in believing that they will get good news. Thus early news provides additional utility to their anticipation. In the real world, over-optimistic people are more likely to search for information than over-pessimistic people. This conclusion contains and further extends KR's conclusion on timing preference of information. KR2009 proved that people will be indifferent to the timing of information, when the sense of loss is exactly as aversive in period 1 as in period 2, that is, $\gamma=1$. Our model repeats their conclusion when agents are rational. However, with biases in beliefs, individuals will have preferences over timing of information even if they have equal sense on prospective and realised gain-loss utilities, and the preferences of early and later information depend on their loss-aversion attitudes, weights on anticipation against realisation and the real chance of a good outcome. 

For the case $\gamma<1$, it is easy to prove that for people holding $q>p$ , we have $U_{early}>U_{wait}$ and early information is good; whereas for $q<p$, $U_{early}$ can be greater, equal to or smaller than $U_{wait}$ for some value of q. Compared with the case in which $\gamma=1$, there is a greater chance that people will prefer early information. KR's explanation is applicable here as the sense of loss for non-immediate outcomes is not as intense, so the agent is better off by receiving the information early. Similar analysis can be applied to $q<p$.

\section{Risk Attitudes}

This chapter explores some significant implications of the most fundamental premise of the model---risk attitudes under subjective beliefs with application to the choice between two independent lotteries. Chapter 3 has analysed the discrete multi-state case. The following study extends the previous conclusions by looking at the case of continuous distributions. The new assumption makes no changes to conclusions in Chapter 3 while it avoids jumps of subjective expectation when moving beliefs from one state to another\footnote{For example, under discrete multi-state case, a small increase in subjective belief of a high rank state may or may not change some other states from gain to loss, while the gain-loss switch always happens to at least one state under continuous distributions.}. To better understand the effects of reference point on risk attitude, this chapter simply assumes that the consumption utility takes the linear format $u(x)=x$, in which case, agents are risk-neutral in the absence of gain-loss utility. The lottery has continuous distribution which can be symmetrical or skewed. Agents are further categorised into two types---\lq\lq naive\rq\rq\ and \lq\lq sophisticated\rq\rq\ based on their different cognitive processes. Both types form their subjective beliefs as previously stated in Chapter 3, however, when gambling, the \lq\lq sophisticated\rq\rq\ type recognises their cognitive biases and make decisions accordingly while the \lq\lq naive\rq\rq\ type fails and behaves like EU maximizer without gain-loss utility. Another rationale for this separation is due to the discussion given by BP2005, in which they explored the behavioural implications of the \lq\lq naive\rq\rq\ type under our definition. In their portfolio choice application, the agent forms optimal subjective beliefs by maximizing the total intertemporal utility, while chooses optimal investment strategy only to maximize current anticipation. One straightforward argument is that agent can choose actions consistent with beliefs in maximizing the total intertemporal utility---leading to the discussion of sophisticated agents in our model\footnote{Consistent actions are chosen simultaneously with beliefs in the sophisticated case. Since there are only two periods in our model, we argue that agents can aim at long-term interests maximization in choosing actions}. Based on the continuity assumption and classification of agents, this chapter explores the risk attitudes implications of the model described in Chapter 3.  

%The lotteries in this part can be viewed as financial assets with short sale constraint binds.
\subsection{Naive Agent}
\begin{ass}\label{a}
$Z_A$ and $Z_B$ are the contingent payoffs of two independent lotteries with continuous probability distribution functions $f_A(\cdot)$ and $f_B(\cdot)$ respectively. 
\end{ass}
\begin{ass}\label{b}
An agent evaluates the payoffs of two lotteries separately and has optimal subjective beliefs $g_A(\cdot)$ and $g_B(\cdot)$ which are solutions to the following problem: 
$$
\underset{g_i(\cdot)}{max}E_{g_i}(Z)+E_{f_i}\mu[Z-E_{g_i}(Z)],
 i=A, B. 
$$
\end{ass}

\begin{ass}{(Naive Agent)}\label{c1}
For any two lotteries A, B satisfying Assumption \ref{a}, a naive agent is the one with optimal beliefs described by Assumption \ref{b} and prefers the lottery with a higher value of: 
$$
\underset{g_i(\cdot)}{Max}E_{g_i(\cdot)}(Z), i=A, B, 
$$
\end{ass}

\begin{prop}{(Two Symmetrically Distributed Lotteries)}\label{twoindep}\\
Suppose the assumptions \ref{a}-\ref{c1} hold for two lotteries $A$ and $B$ satisfying: 
(i) $E_{f_A(\cdot)}(Z)=E_{f_B(\cdot)}(Z)$, for $f_A(\cdot)\neq f_B(\cdot)$; \\
(ii) $Z_A$ and $Z_B$ are both symmetrically distributed;\\
(iii) $Z_A$ and $Z_B$ satisfy the single-crossing property, that is, if $F_A(\cdot)$ and $F_B(\cdot)$ are the cumulative distribution functions for $Z_A$ and $Z_B$, there exists $z$ such that $F_A(x)<F_B(x)$ for $x<z$ and $F_A(x)>F_B(x)$ for $x>z$. \\
Then,\\
an agent with a cut-off probability $P^*\geq 1-F(z)$ prefers lottery A to B since $E_{g_A(\cdot)}(Z)>E_{g_B(\cdot)}(Z)$;\\
an agent with a cut-off probability $P^* \leq 1-F(z)$ prefers lottery B to A since $E_{g_B(\cdot)}(Z)>E_{g_A(\cdot)}(Z)$;
\end{prop}

Proposition \ref{twoindep} says that a naive agent with optimal beliefs is 
%a risk-neutral\footnote{An agent with linear utility function is defined as risk-neutral in neoclassical economics.}subjective expectation maximizer is 
risk-seeking if optimistic and risk-averse if pessimistic
%\footnote{The terms \lq\lq risk-seeking\rq\rq\ and \lq\lq risk-averse\rq\rq\ have been well-defined in neoclassical economics, but here we use them to describe people's general tastes for lotteries. }.
Intuitively, low $P^*$ implies weak loss-aversion and less valued future.  Loss in the future is therefore bearable in this case. Higher risk lottery provides better chances of gains on the right tail of the distribution, which is treated as stronger evidence of a promising payoff by an optimistic agent---leading to further up-biases. The positive prospect from over-estimated chances of gains can significantly increase their anticipatory utility, while the negative prospect from more painful feeling of loss are countered by the further underestimation of probabilities in the bad outcomes. 

For symmetrical distributions with single-crossing property, the risky lottery gives a higher cumulative probability beyond the crossing point, meaning that the risky assets promise a higher chance of gain over a certain level of expectation. Therefore, a naive agent with a low cut-off $P^*$ is more up-biased in lotteries with fat tails. 
%Financial assets can be viewed as lotteries with short-sale constraint binds. 

This proposition implies that a mean-preserved spreading is desirable for an optimistic agent and undesirable for a pessimistic one even if agents are risk-neutral under traditional economics definition. 

%The intuition behind this conclusion is that the mean-preserved spread fattens the right tail of the wealth distribution--generated more chances of gains--but also fattens the left tail of the distribution--generated more chances of losses. As investors with different cut-off points have biased beliefs and therefore different levels of expectations, the mean-preserved spreading would have unsymmetrical effects on gains and losses even though the real wealth distribution is symmetric. Specifically, a mean-preserved spread gives more extra gains than extra losses to an investor with low expectation while gives more extra losses than gains to an investor with high expectation. 

%is over-optimistic on both assets' returns. Both subjective expectations are higher than the unbiased level. Since the total probability beyond the crossing point is higher for the risker asset, a 
Based on the proof of Proposition \ref{twoindep} and the intuitions described above, we derive the following lemmas:

\begin{lemma}{(Optimistic and pessimistic risk attitude)}\\
For the group of symmetric distributions, a mean-preserving spreading is desirable for an optimistic agent and undesirable for a pessimistic agent. 
\end{lemma}
\begin{lemma}{(Ranking of subjective expectations)}\\
For any distribution, subjective expectation is non-increasing in $P^*$. 
\end{lemma}
Furthermore, we consider a more general version of Proposition \ref {twoindep} by relaxing the requirement on symmetry. 
\begin{prop}{(Two Lotteries: The General Case)}\label{twoindepgen}\\
Consider two independent lotteries $A$ and $B$ with continuous and differentiable distribution functions $f_A(\cdot)\neq f_B(\cdot)$ and $E_{f_A(\cdot)}(Z)=E_{f_B(\cdot)}(Z)$. For an agent with cut-off probability $P^*=\dfrac{\eta \lambda-1}{\eta(\lambda-1)}$, we have:
\begin{itemize}
\item[(i)]If $P_{+A}^{0}>(<)P^{*},P_{+B}^{0}<(>)P^{*}$, then \\
 $E_{g_{A}^{*}}(Z)>E_{g_{B}^{*}}(Z)$ ($E_{g_{A}^{*}}(Z)<E_{g_{B}^{*}}(Z)$);
\item[(ii)]If $P_{+A}^{0}>(<)P^{*},P_{+B}^{0}>(<)P^{*}$, then\\
 $E_{g_{A}^{*}}(Z)>E_{g_{B}^{*}}(Z)$ ($E_{g_{A}^{*}}(Z)<E_{g_{B}^{*}}(Z)$)\\
iff $\underset{a}{\overset{+\infty}{\int}}[f_{A}(Z)-f_{B}(Z)]\mathrm{d}Z>0$ ( $\underset{b}{\overset{+\infty}{\int}}[f_{B}(Z)-f_{A}(Z)]\mathrm{d}Z>0$), 
\end{itemize}
where $a, b, P_{+A}^{0}, P_{+B}^{0}$ are\\
$P^{*}=\underset{a}{\overset{+\infty}{\int}}f_{A}(Z)\mathrm{d}Z=\underset{b}{\overset{+\infty}{\int}}f_{B}(Z)dZ$, $P_{+A}^{0}=\overset{+\infty}{\underset{E_{f_{A}}}{\int}}f_{A}(Z)\mathrm{d}Z$,  $P_{+B}^{0}=\overset{+\infty}{\underset{E_{f_{B}}}{\int}}f_{B}(Z)\mathrm{d}Z$.
\end{prop}

Notice that $a$ and $b$ in Proposition \ref{twoindepgen} are the subjective expectations under optimal beliefs. Since optimal expectations are set at the level to ensure the cumulative probability above it equals $P^*$, $a$ and $b$ are also indicators of the distribution skewness. For any given $P^*$, a higher value of $a$ means a fatter \lq\lq right tail\rq\rq\ of the distribution and therefore a negative skewness\footnote{The \lq\lq right tail\rq\rq\ does not only refer to the tail of a distribution. It represents the area under the p.d.f. from the subjective expectation controlled by $P^*$ to the right limit.}. As long as the mean remains the same, an optimistic agent has a decreasing preference to lotteries as the skewness of distribution changes from negative to positive. 

\subsection{Sophisticated Agent}

\begin{ass}{(Sophisticated Agent)}\label{c2} 
For any two lotteries A and B satisfying Assumption \ref{a}, a sophisticated agent is the one with optimal beliefs described by Assumption \ref{b} and prefers the lottery with higher value of 
$$
\underset{g_i(\cdot)}{Max}E_{g_i}(Z)+E_{f_i}\mu[Z-E_{g_i}(Z)], i=A, B, 
$$
\end{ass}
\begin{prop}{(Choice between Two Lotteries: Sophisticated Case)}\label{lotterys}\\
If assumptions \ref{a},\ref{b}, and \ref{c2} hold, then for any two risky lotteries with  $f_A(\cdot)\neq f_B(\cdot)$, a sophisticated agent strictly prefers lottery A to B iff
$$
\int_{E_{g_A}}^{+\infty}f_A(Z)Z\mathrm{d}Z+\lambda\int_{-\infty}^{E_{g_A}}f_A(Z)Z\mathrm{d}Z>\int_{E_{g_B}}^{+\infty}f_B(Z)Z\mathrm{d}Z+\lambda\int_{-\infty}^{E_{g_B}}f_B(Z)Z\mathrm{d}Z
$$
Specifically, for two risky lotteries with equal objective expectation, i.e., $E_{f_A(\cdot)}(Z)=E_{f_B(\cdot)}(Z)$, a sophisticated agent prefers lottery A to B iff,
$$
\int_{-\infty}^{E_{g_A}}f_A(Z)Z\mathrm{d}Z>\int_{-\infty}^{E_{g_B}}f_B(Z)Z\mathrm{d}Z
$$
\end{prop}

As we can see from Proposition \ref{lotterys}, the objective function of a sophisticated agent is just the expectation under rational beliefs but weighting the loss region with $\lambda$. Without loss aversion, i.e., $\lambda=1$, the objective function of an sophisticated agent is simplified to the ordinary rational expectation. 

Proposition \ref{lotterys} indicates that a sophisticated agent behaves similarly to a rational agent who maximizes her expectation with unbiased beliefs. There is no surprising for this conclusion since a sophisticated agent considers both the anticipation and the deterioration in realization from the \lq\lq reference effect\rq\rq\ . Here, the \lq\lq reference effect\rq\rq\ denotes the deduction in realized utility coming from comparing an outcome with the anticipation(reference point). We notice that the direct effects from subjective beliefs, i.e., anticipatory utility, is eliminated by the \lq\lq reference effect\rq\rq\ at optimal beliefs as we prove in Appendix A. Intuitively, if the anticipation is too high, then the \lq\lq reference effect\rq\rq\ gives too many chances of loss. Since the agent is loss averse, decrease in total utility due to more states of loss exceeds the increase in anticipatory utility. Therefore, lower anticipation is beneficial. On the other hand, with low anticipation, few chances of loss makes the punishment from the \lq\lq reference effect\rq\rq\ insignificant. Since people discount the future utility, benefits from higher anticipation will exceed the decrease in future realised utility. At optimal beliefs, illusions from subjective beliefs on anticipation and reference point cancel out each other, leaving only the truth of the contingency. 

Our objective function here deviates from the rational expectation only in the loss region. This difference comes directly from the assumption of loss aversion since loss averse agent overweighs the bad outcomes below their expectation. 

\section{Application: Portfolio Choice}
%Investors with high expecation are more sensitive to the realization of result, since they have larger chances of losses. Since people are loss averse, therefore, providing more fluctuations in utility. 
%However, whether higher sensitivity is good or bad is determined by the real distribution of the underlying asset. For sophisticated ones, who excluded the effects from anticipatory utility, still could not avoid their sensitivity to the results in loss region. People are always sensitive to the lower ranking returns than to the higher ranking returns, determined by the loss aversion. Therefore, 
%They have in general stronger feelings of results. when the returns are good, then, stronger feelings are better, therefore, take more risks. Otherwise, take less risks. 
%Great!! Finally got it!

This chapter explores some implications of the model in the investment problem---a biased investor with concave utility function chooses the optimal portfolio consisting of a risky asset and a risk-free asset. Following the assumption made in Chapter 4, we assume that there exist two types of investors: the naive ones who maximize the expected return and the sophisticated ones who maximize the total utility including both the anticipatory and prospective gain-loss utilities. We points out that the naive investor and the sophisticated investor can adopt opposite investment strategies even though they form their subjective beliefs through the same cognitive process. An equilibrium is further derived in a market with identical investors to look into the implications  on asset pricing. 
\subsection{Choice between one risk-free asset and one risky asset}\label{riskfree}
\begin{ass}
The return of an asset has a continuous distribution function $f(\cdot)$ defined over $(-\infty,+\infty)$. The investor holds subjective belief $g(\cdot)$ of the asset's return, which is also a continuous $p.d.f.$ on $(-\infty,+\infty)$. 
\end{ass}
There are two assets in the market: a risk-free asset with return $R_f$; and a risky asset with gross return $R_R=R_f+R_m=R_f+E_{f(\cdot)}(R)$, where $R_m$ is the gross excess return, $R$ the realised excess return. Investors have unlimited access to the risk-free asset and the price of the risk-free asset is 0. The consumption utility takes the same assumption as in Chapter 3: $u'(x)>0, u''(x)\leq0$. There are two periods. At t=1, the agent forms her subjective optimal beliefs $g(R)$ about payoffs of the risky asset and allocates her unit endowment between these two assets. In the second period, the payoffs of the assets are realised. 

Following our previous categorisation of \lq\lq naive\rq\rq\ and \lq\lq sophisticated\rq\rq\ agents, we assume that the naive type has different objective functions in choosing optimal beliefs and optimal $\alpha$. The choice of portfolio is \lq\lq rational\rq\rq\ based on their biased beliefs. The sophisticated type chooses optimal subjective beliefs and optimal $\alpha$ allocated to the risky asset simultaneously. 

To be more specific, at t=1, for any given optimal beliefs $g(R)$, the naive agent chooses her portfolio share, $\alpha^{BS}$ allocated to the risky asset to maximize the expected return: 
\begin{equation*}
\underset{\alpha}{Max}\int_{-\infty}^{+\infty}g(R)u(R_{f}+\alpha R)\mathrm{d}R. 
\end{equation*}
Given the optimal choice of $\alpha^{BS}$, the naive agent chooses her subjective beliefs $g(\cdot)$ to solve: 
\begin{equation*}
\underset{g(R)}{argmax}\: E_{g(R)}u(R_{f}+\alpha^{BS}R)+E_{f(R)}\mu[u(R_{f}+\alpha^{BS}R)-E_{g(R)}u(R_{f}+\alpha^{BS}R)] 
\end{equation*}
Instead, a sophisticated agent chooses her optimal portfolio share $\alpha$ and optimal beliefs $g(R)$ simultaneously to solve: 
\begin{equation*}
\underset{g(R),\: \alpha}{argmax}\: E_{g(R)}u(R_{f}+\alpha R)+E_{f(R)}\mu[u(R_{f}+\alpha R)-E_{g(R)}u(R_{f}+\alpha R)] 
\end{equation*}

\cite{Brun05} built a discrete model in which an agent has optimal beliefs determined by \\
\begin{equation*}
\underset{\{q_{s}\}}{argmax}\:\overset{S}{\underset{s=1}{\Sigma}}q_{s}u(R_{f}+\alpha^{BS}R_{s})+\overset{S}{\underset{s=1}{\Sigma}}p_{s}u(R_{f}+\alpha^{BS}R),
\end{equation*}
where $\alpha^{BS}$ is the solution to $\underset{\alpha}{max}\:\overset{S}{\underset{s=1}{\Sigma}}q_{s}u(R_{f}+\alpha R_{s})$. 
This is a special case of the naive agent in our model when $\lambda=1$ and $\eta=\dfrac{1}{2}$. 

BP2005 concluded that optimal-belief holders  always trade more aggressively than rational agents since the cost from distorted portfolio choice is second order while the benefit from a higher anticipation is first order. 

The following proposition indicates that reference-dependent loss-averse investors no longer trade aggressively all the time. Even though the cost of distorted portfolio is still second order, biased beliefs not only generate first-order benefits from higher anticipation, but also give first-order punishment from the \lq\lq reference effect\rq\rq\ . Therefore, agents in our model take aggressive or conservative trading strategy depending on their optimistic and pessimistic attitudes.

\subsubsection{Naive Agent}

\begin{prop}{(Risk Taking due to Optimism and Pessimism: Naive Case):}\label{riskfreenaive}\\
An optimistic investor with low $P^*$ invests more aggressively than a rational investor or in the opposite direction; a pessimistic investor with high $P^*$ invests in the same direction  as the rational investor but more conservatively: \\
if $E(R)>0$, then $\alpha^{RE}>0$, 
$$
\alpha^{OP}>\alpha^{RE}>0 \; or\; \alpha^{OP}<0<\alpha^{RE}; \;
0<\alpha^{PE}<\alpha^{RE}. 
$$
If $E(R)<0$, then $\alpha^{RE}<0$,
$$ \alpha^{OP}<\alpha^{RE}<0 \; or\; \alpha^{OP}>0>\alpha^{RE}; \;
\alpha^{RE}<\alpha^{PE}<0.
$$
\end{prop}

Similar to BP2005, the optimal $\alpha^{BS}$ under biased beliefs is always different from $\alpha^{RE}$ since biased beliefs ensure higher total welfare as proved in Proposition 1. Besides, the optimistic investor trades in the same direction but more aggressively than a rational agent since she overestimates the chance of good returns\footnote{Returns are good or bad conditional on the long-short position. A positive return is good conditional on long and is bad conditional on short.}; or she enters into a position opposite to the rational strategy. As stated at the beginning of this chapter, the investor in BP's model can be described as an optimistic agent with $P^*=0$ in our model and the behavioural implication on an optimistic agent here can also be explained by BP's intuition: opposite trading happens when the asset is skewed enough in the opposite direction of the mean payoff. 

Different from BP's conclusion, the investor in our model no longer always holds \lq\lq optimistic\rq\rq\  beliefs and invests aggressively. Instead, a pessimistic investor characterised  by a high $P^*$ invests in the same direction but more conservatively than her rational counterpart\footnote{\lq\lq Optimistic\rq\rq\ and \lq\lq pessimistic\rq\rq\  are defined differently from \cite{Brun05}. BP's \lq\lq pessimistic\rq\rq\ investor assigns higher probabilities to negative returns while shorting the asset. \lq\lq Pessimistic\rq\rq\ defined in this model means \lq\lq conservative\rq\rq\  : pessimistic investors assign lower probabilities to negative returns while still shorting.}. Agents in BP's model always trade aggressively because the additional anticipation is a pure generator of felicity\footnote{Mathematically, BP's model requires $\alpha^{BS}>\alpha^{RE}>0$ because their first order maximization problem requires this condition. The envelope condition is $(u_{s''}-u_{s'})\Delta\hat{\pi}+\eta\Sigma p_{s}u'(R_{f}+\alpha^{BS}R_{s})R_{s}\mathrm{\Delta\alpha}=0$. When $\alpha^{RE}>\alpha^{BS}>0$,  $\Sigma p_{s}u'(R_{f}+\alpha^{BS}R_{s})R_{s}>0$, then bias in beliefs serves as a utility pump thus gives no limits to bias. }. Although optimism and pessimism are both punished for distorted portfolio choices, the former dominates the latter because up-biases provide extra felicity from the higher anticipation whereas down-biases do not. 

Moreover, an agent with a low $P^*$ in our model tends to overestimate the chance of good returns\footnote{Whether a return is good or bad is conditional on the position of the asset, i.e., a positive return is good conditional on long and bad conditional on short.} no matter they have a short or long position. However, attitudes can also be strategy-dependent---the same person in a long position can exhibit an attitude opposite to that when she shorts. For example, facing a negative skewed asset with $E(R)<0$, an investor enters into a short position will be pessimistic since the long tail on the left gives few chances of gains; the same investor in the long position instead can be optimistic, since the fat tail on the right gives high chance of gain. The choice of attitudes depends on the investment position as well as the skewness of the risky asset. 

Finally, we point out that our conclusion partially depends on the construction of the optimization problem. By assuming that the agent takes beliefs as given when making portfolio choice and fails to recognise her biases, we implicitly indicate, in a non-strict way,  that the agent in our model chooses $\alpha$ after she chooses her beliefs. Therefore, the deviation of $\alpha$ from rational level, like BP said, only puts on second order cost whereas changes in beliefs introduce first order increase in anticipation hence dominate the total effects. However, an agent can behave more complicatedly than the described naive type---when choosing the optimal portfolio, she realises her biases in beliefs and makes her investment decision accordingly. In the following part, we discuss the behaviour of a sophisticated agent in the portfolio choice problem.

\subsubsection{Sophisticated Agent}

\begin{prop}{(Risk Taking due to Optimism and Pessimism: Sophisticated Case):}\label{riskfreeso}
\begin{itemize}
\item[(i)] 
Sophisticated investors invest more aggressively than rational investors if the average return conditional on loss is good enough. In other cases, sophisticated investors invest either more conservatively or in the opposite direction to the rational strategy.\\
If $E(R)>(<)0$ and $\underset{-BS}{\int}f(R)u'(R_{f}+\alpha_{RE}R)R\mathrm{dR}>(<)0 $, then,
$$
\alpha^{BS}>\alpha^{RE}>0 (\alpha^{BS}<\alpha^{RE}<0). 
$$
If $E(R)>(<)0$ and $\underset{-BS}{\int}f(R)u'(R_{f}+\alpha_{RE}R)R\mathrm{dR}<(>)0 $, then
$$
\alpha^{BS}<\alpha^{RE} (\alpha^{BS}>\alpha^{RE})
$$
where $\alpha^{BS}$ and $\alpha^{RE}$ are the optimal allocation of wealth on risky assets and \lq\lq -BS\rq\rq\  indicates the region of loss under biased beliefs taking $\alpha_{RE}$ as given.
\item[(ii)]
$\alpha^{BS}$ decreases in $P^*$ if $R_{CE}>0$;\\ 
$\alpha^{BS}$ increases in $P^*$ if $R_{CE}<0$, \\
where $E_{g(\cdot)}u(R_f+\alpha R)=u(R_f+\alpha R_{CE})$ and $R_f+R_{CE}$ is the certainty equivalent . 
\end{itemize}
\end{prop}
%Optimistic agents invest more aggressively than pessimistic agents if the return of the assets is perceived as good on average;
%Pessimistic agents invest more aggressively than optimistic agents if the return of the asset is perceived as bad on average. 
%Formally, suppose $E_{g(\cdot)}u(R_f+\alpha R)=u(R_f+\alpha R_{CE})$, where $R_f+R_{CE}$ is the certainty equivalent and $E_{g(\cdot)}u(R_f+\alpha R)$ is the expectation under optimal subjective beliefs. 
%For $R_{CE}\geq0$, we have, 
%$$
%0<\alpha^{PE}<\alpha^{OP} or \alpha^{PE}<\alpha^{OP}<0 
%$$
%For $R_{CE}<0$, we have, 
%$$
%\alpha^{OP}<\alpha^{PE}<0 or \alpha^{PE}>\alpha^{OP}>0
%$$
%where $\alpha^{OP}$ and  $\alpha^{PE}$ are the proportion of wealth allocated to risky asset with optimal optimistic and pessimistic beliefs respectively. 
%\end{prop}

%For $\alpha^{RE}\geq0(<0)$ and $R_{CE}\geq0$, we have, 
%$$
%\alpha^{PE}<\alpha^{RE}<\alpha^{OP}  (\alpha^{OP}<\alpha^{RE}<\alpha^{PE});
%$$
%For $\alpha^{RE}\geq0(<0)$ and $R_{CE}<0$, we have, 
%\alpha^{OP}<\alpha^{RE}<\alpha^{PE} (\alpha^{PE}<\alpha^{RE}<\alpha^{OP});where $\alpha^{RE}$ is the optimal wealth proportion allocated to the risky asset under rational unbiased beliefs, $\alpha^{OP}$ and  $\alpha^{PE}$ are the proportion to risky asset with optimal optimistic and pessimistic beliefs respectively. \end{prop}

The behaviour of sophisticated agents is more complicated as described by proposition \ref{riskfreeso}. Like our analysis in Chapter 4, a sophisticated agent considers both anticipation and the \lq\lq reference effect\rq\rq\  in making her portfolio choice. At optimal beliefs, these two effects cancel out each other, leaving only rational-like expected utility. Unlike naive agents, the biased beliefs no longer determine actions directly through expectation, but they still affect actions through the choice of reference point  which determines the sensitivity towards future payoffs. Basically, optimism, by introducing in more losses, also makes an agent more sensitive towards outcomes while pessimism, by reducing the fear of loss, makes an agent less sensitive to outcomes. 

The first part of Proposition \ref{riskfreeso} describes the difference in portfolio choice strategies between a sophisticated and a rational agent, which lies in whether they overweigh  the returns in loss region. The intuition is simple: compared with a rational agent, a loss-averse agent cares more about low-rank outcomes. Therefore, when low-rank returns are good on average, loss-averse  agents are happier than the rational ones and more willing to take excess risks. On the other hand, bad returns on the low-rank outcomes are more painful for loss-averse  agents thus lead  to conservative or opposite trading strategy.  

The second part of Proposition \ref{riskfreeso} examines the effects of optimistic and pessimistic attitudes on investment strategies. To understand the intuition more clearly, consider the following lemma directly derived from Proposition \ref{riskfreeso} part (ii).  
\begin{lemma}\label{riskfreesolemma}
Optimistic agents invest more aggressively than pessimistic agents when the return is perceived to be good;\\
Pessimistic agents invest more aggressively than optimistic agents when the return is perceived to be bad. That is, \\
for $\alpha^{BS}>(<)0$ and $R_{CE}>0$, we have,
$$
\alpha^{OP}>\alpha^{PE}>0 (\alpha^{PE}<\alpha^{OP}<0);
$$
for $\alpha^{BS}>(<)0$ and $R_{CE}<0$, we have,
$$
\alpha^{PE}>\alpha^{OP}>0 (\alpha^{OP}<\alpha^{PE}<0).
$$
\end{lemma}

To better understand the intuition, consider the following example. Suppose there is an investor in the long position of a security and suppose the security gives good returns on average in the future which is known by the investor. An optimistic investor who takes her large chance of future loss into consideration cares more about the returns due to her fear of loss. Since loss is the source of intensive feelings compared to gains, more attention to the asset gives stronger feeling on returns. Hence, when returns are good  in general, an optimistic investor is happier than a pessimistic one and is willing to hold more risky asset. 
%Even though a pessimistic sophisticated investor knows that her low anticipation can avoid painful loss in the future, however, being pessimistic also makes her numb since she has less fear for loss and thus less care for any outcome. We say that she is numb because loss is source of strong feelings compared to gains. By avoiding losses, she also avoids intensive feelings. For this reason, good future prospects provides less felicity. Instead, 
 
On the contrary, pessimism reduces the fear for loss and makes the agent less sensitive, thus care less about future outcomes . Therefore, when the asset is a bad one, pessimism and numbness make the bad outcomes more tolerable and the pessimistic agents can bear extra risks than optimistic ones. 

Compared with the bad-return situation, when the asset gives good return as in the first case, even though staying numb can make pessimistic ones avoid intensive feelings from bad outcomes, there is little chance to feel bad since the asset is a good one in general. Therefore, optimism is more beneficial because it intensifies the happiness and enables optimistic investors to bear more risks than pessimistic ones.
%Changing beliefs would change the value of expectation, therefore, enclose more or less return levels into the "gain" region. However, the marginal utility by doing so is $u'(R_f+R)R$, depends on whether the return is positive or negative. An increase in $\alpha$ leads to a second order decrease in utility that is going to be realized as marginal utility is decreasing: $u"(\cdot)\leq0$. Optimal beliefs must further change to compensate the decrease in utility. When the marginal utility is positive, then further up-bias is beneficial since it gives a higher first-order utility as $u'(\cdot)\geq0$. Therefore, if an investor is optimistic at her optimal beliefs, then, bring back her beliefs to the rational level would give her a first order decrease in anticipatory utility. Therefore, the investor must choose a lower $\alpha$ to increase the marginal utility. Instead, when the marginal certainty equivalent is negative, bring down the beliefs would lead to a first order increase in marginal utility. The investor wants to hold more risky assets to increase the total utility. Similar analysis for pessimistic investors. Compared to naive agents, optimistic investors are no longer simply holding more risky assets as they over evaluate its the return and pessimistic investors are no longer just shorting the risky assets, instead, both optimistic and pessimistic investors can hold more or less proportion of the risky assets than their rational counterpart. 
\subsection{Equilibrium}
In this section, we place the portfolio choice problem in an exchange economy with two assets, a free risk-free one with $R_f=0$ and price $\pi_f=0$ and a risky one with stochastic excess return $R$ and price $\pi$. The distribution of $R$ is publicly known. We assume the short-sale constraint binds, thus the proportion of wealth allocated to the risky asset $\alpha$ satisfies $0 \leq \alpha \leq 1$. We further simplify the assumption by making utility function take the linear format: $u(x)=x$. 

The simplest candidate equilibrium is a homogeneous holdings equilibrium: an equilibrium in which investors are identical to each other and hold the same portfolio. As we prove later, the optimal portfolios for both naive and sophisticated agents are either only risky asset or only risk-free asset  under the linear utility function assumption. Hence, the equilibrium price which is constructed by such kind of portfolio choice must ensure investors are indifferent between two assets.

\subsubsection{Naive}
Previous problem of a  naive agent under the new assumption becomes: 
$$
g^{*}(R)=\underset{g(R)}{argmax}\underset{-\infty}{\overset{+\infty}{\int}}g(R)(R_{f}+\alpha^{*}R)\mathrm{dR}+\underset{-\infty}{\overset{+\infty}{\int}}f(R)[R_{f}+\alpha^{*}R-\underset{-\infty}{\overset{+\infty}{\int}}g(r)(R_{f}+\alpha^{*}r)\mathrm{dr}]\mathrm{dR}
$$ 
\begin{equation*}
\alpha^{*}=\underset{\alpha\in[0,1]}{argmax}\underset{-\infty}{\overset{+\infty}{\int}}g^{*}(R)(R_{f}+\alpha R)\mathrm{dR}, 
\end{equation*}\\
It is obvious from the optimization problem that for $R_f=0$, we have,
\begin{center}
$\alpha^*=1$ if $\quad E_{g^{*}}(R)\geq 0$;\\
$\alpha^*=0$ if $\quad E_{g^{*}}(R)\leq 0$.
\end{center}
In equilibrium, the price of the risky asset $\pi_{N}$ at $t=1$ must satisfy $R_{f}=\eta E_{g^{*}} (R)-\pi_{N}$. 
The equilibrium price is therefore, 
$$
\pi^{*}_{N}=\eta E_{g^{*}}(R) 
$$
Instead, in a market with rational unbiased EU maximizers, the equilibrium price of the risky asset at t=1 is,
$$\pi^{*}_{RE}=\eta E_{f}(R).$$

\subsubsection{Sophisticated}
Now consider the security pricing if investors are sophisticated. The optimization problem is defined by,
$$
\underset{g(R),\alpha}{max}\underset{-\infty}{\overset{+\infty}{\int}}g(R)[R_{f}+\alpha R]\mathrm{dR}+\underset{-\infty}{\overset{+\infty}{\int}}f(R)\{R_{f}+\alpha R-\underset{-\infty}{\overset{+\infty}{\int}}g(r)[R_{f}+\alpha r\mathrm{dr}\}\mathrm{dR}
$$
From the first order condition with respect to $\alpha$, we have, 
\begin{center}
$\alpha=1$ if $\underset{-\infty}{\overset{+\infty}{\int}}f(R)R\mathrm{dR}+(\lambda-1)\int_{-\infty}^{E_{g^{*}}}f(R)R\mathrm{dR}-\pi_{S}>R_f$;\\
$\alpha=0$ if $\underset{-\infty}{\overset{+\infty}{\int}}f(R)R\mathrm{dR}+(\lambda-1)\int_{-\infty}^{E_{g^{*}}}f(R)R\mathrm{dR}-\pi_{S}<R_f$.
\end{center}
In equilibrium, the price of the risky asset $\pi_S$ must satisfy
\begin{equation*}
\underset{-\infty}{\overset{+\infty}{\int}}f(R)R\mathrm{dR}+(\lambda-1)\int_{-\infty}^{E_{g^{*}}}f(R)R\mathrm{dR}-\pi_{S}=R_f=0
\end{equation*}
Therefore, the equilibrium price for a market with identical sophisticated agents is \\
\begin{equation*}
\pi_{S}^{*}=\eta E_f(R)+\eta(\lambda-1)\int_{-\infty}^{E_{g^{*}}}f(R)R\mathrm{dR}. 
\end{equation*}
%Instead, if investors are rational holding unbiased beliefs, we have
%\begin{equation*}
%\pi_{SR}^{*}=E_{f}(R)+\eta(\lambda-1)\int_{-\infty}^{E_{f}}f(R)R\mathrm{dR}-\eta(\lambda-1)E_{f}(R)\cdot\int_{-\infty}^{E_{f}}f(R)\mathrm{dR}
%\end{equation*}. 

\begin{figure}
\centering
\includegraphics[scale= 0.065]{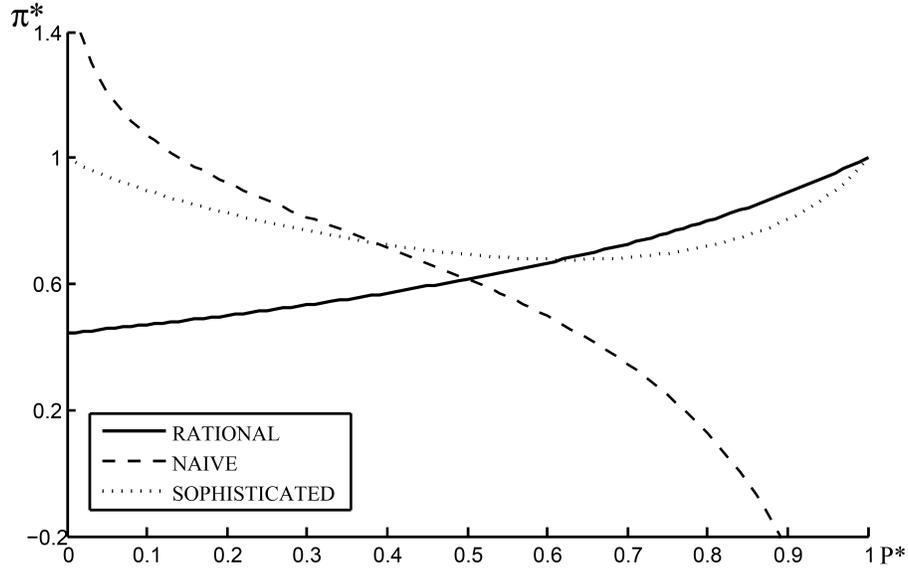}
\caption{EQUILIBRIUM PRICE $\pi^*$ AS A FUNCTION OF $P^*$}
\label{eq}
\end{figure}

To gain further intuition on the equilibrium, consider a normal distributed asset with $R\sim N(1, 1)$.  $\lambda$ is set to $2.25$ as suggested by experimental evidence . Variation in  $\eta$ changes $P^*$ from $0$ to $1$\footnote{$\lambda$ and $\eta$ can change simultaneously, however, for calibration purpose, we set lambda at 2.25.}.  The result is shown in Figure\ref{eq}.

The solid line in Figure \ref{eq} plots the rational price as a function of  $P^{*}$. Since $P^{*}$ is increasing in $\eta$, the equilibrium price increases to rational expectation as the discount rate $\eta$ goes to 1. The dash line plots the equilibrium price against $P^{*}$ in a market with naive investors. Consistent with the prediction of Proposition \ref{riskfreenaive} and since the short-sale constraint binds , the equilibrium price of the risky asset decreases as investors become more pessimistic. To be more specific, the \lq\lq naive\rq\rq\ price is higher than the rational level when the market is dominated by optimistic investors because the up-biased beliefs urge the investors to hold more risky asset. Instead, the \lq\lq naive\rq\rq\  price falls below rational level when investors are pessimistic. Pessimistic investors require higher equity premium to compensate for their stronger aversion to loss.

%To gain further intuition on the equilibrium, consider the following example with previous discussed assets: a normal distributed asset with $R\sim N(1, 1)$ and a positive skewed asset with $R \sim Gamma(4, 0.5)$. Under these assumptions, these two assets share the same expectation and variance. Furthermore, $\lambda$ is set at $2.25$ as suggested by experimental evidence. Variation in $\eta$ changes $P^*$ from $0$ to $1$. The result is shown in Figure X and X. 

The dotted line plots the \lq\lq sophisticated\rq\rq\ equilibrium price as a function of $P^*$. Compared with the \lq\lq naive\rq\rq\ market, the sophisticated price exhibits  a U-shaped patten rather than strictly decreasing in $P^*$. The equilibrium price is upper bounded by the rational expectation reflecting our conclusion that sophisticated agents act similarly to rational agents. Moreover, consistent with the prediction of Proposition\ref{riskfreeso}, both the optimistic and the \lq\lq very\rq\rq\ pessimistic agents price the asset higher than moderate pessimistic agents. With short-sale constraint, optimistic agents for this symmetrically distributed asset are those with $P^{*}<0.5$ while pessimistic agents have $P^{*}>0.5$. Since $\mu=1$, $R_{CE}^{OP}=E_g(R)>0$ and utility increases as agents become more optimistic, so does the price. The average return captured by the optimistic agents is always greater than 0, therefore, the asset is a good one and the higher reference point intensify the happiness for good returns.  

On the contrary, for \lq\lq extremely\rq\rq\ pessimistic agents (in this case, $P^{*}>0.63$), they have $R_{CE}^{PE}=E_g(R)<0$ and the risky asset becomes more appealing as agents turn to be more pessimistic. As the return of the asset is not good enough for those very pessimistic agents to bias up, it is subjectively captured to be bad. Further pessimism reduces sensitivity toward bad returns and increases tolerance of risks. 

Finally, \lq\lq moderately\rq\rq\ pessimistic agents who have $0.5<P^{*}<0.63$ in this example, correctly  capture the goodness of the asset, that is they have $R_{CE}^{PE}=E_g(R)>0$. They are pessimistic because the asset fails to give sufficient chance of good returns to persuade them to be optimistic. By being more pessimistic, they reduce the deserved happiness from good outcomes instead of reducing the painfulness from bad outcomes. Therefore, they are the most \lq\lq risk-averse\rq\rq\ ones among these three groups. 

Furthermore, rational price is lower than \lq\lq sophisticated\rq\rq\ price when investors are optimistic and exceeds the latter as investors become more pessimistic. Consistent with the first part of Proposition \ref{riskfreeso}, since $\mu>0$, the average return of  the loss region increases from negative to positive as confidence increases. Consequently, intensive feelings on bad returns push down the sophisticated price when market is pessimistic and boost it up when the market is optimistic.

\section{Conclusion}
This thesis develops a model of optimal judgemental biases with reference point. The model gives an insight into the rationale of over-optimism and over-pessimism by applying two behavioural assumptions---reference dependent utility and loss aversion into an intertemporal model. Contrary to previous literatures on reference dependent utility models in which it is usually assumed that utility is derived from beliefs, in our model, beliefs are optimally determined by the inverse process through utility maximization. Our model setting internalizes the over and under confidence commonly observed without employing the ad hoc parameter controlling optimistic and pessimistic attitudes. Applying the model to the preference of information timing sheds lights on the information-seeking behaviour of individuals holding biased beliefs; another example in the context of portfolio choice shows that pessimism can lead to conservative trading but can also encourage risk-taking investment strategy. 

As part of the optimal beliefs literature, the model  we built in this thesis is still open to the critiques of \cite{Spiegler08} on the violation of IIA, since beliefs are payoff dependent. Further amendments might require preferences to be choice-set  dependent---the same element in different choice bundles need to be viewed as different subjects. Moreover, the set of optimal beliefs in our model is not unique. Although the results are compatible  with cumulative prospect theory and other theories on ambiguity, our model cannot  provide explanations to phenomenon like \lq\lq curse of knowledge\rq\rq\\footnote{The curse of knowledge is a cognitive bias according to which better-informed agents may have the disadvantage that they lose some ability to understand less-informed agents. As such added information may convey some disutility, the curse of knowledge implies that the well-informed party, in our model, with a great chance of getting good results, are more likely to be pessimistic instead of optimistic as we predicted.}. One possible solution to this problem is to further assume a reference-dependent anticipation. Reference-dependent anticipation assumption can be valid in the circumstances when agents have fresh recently experience \footnote{Another example to support this assumption is an amateur and a professional chess players having different levels of felicities from the same anticipation.}. Calibration results indicate that with concave utility function, model with reference point may better fit the observations. 

Future work may include modelling of preference on information containing uncertainty. Agents in our model are motivated to choose subjective beliefs deviated from rational ones due to the fear of loss and the appeal of goodwill. Biased beliefs regarding information containing uncertainty is one step further. However, Bayes rule can hardly hold here since beliefs are utility-serving. A simple rule of updating beliefs with economics intuitions is worth exploring. Finally, the equilibrium we describe in this  thesis is incomplete. A more general equilibrium can be constructed based on both our current homogeneous assumption as well as an alternative one that investors holding heterogeneous beliefs trade with each other in the market.

% The appendix command is issued once, prior to all appendices, if any.
\appendix

\section{Mathematical Appendix A}
\textbf{Proposition \ref{bsbetter}: (Biased Beliefs are Preferred to Rational Unbiased Beliefs)\\}
$$U=\underset{s\in\mathbf{\mathcal{S}}}{\sum}q_{s}u_{s}+\eta\{\underset{s\in\mathbf{\mathcal{S}}}{\sum}p_{s}\mu(u_{s}-\underset{s\in\mathbf{\mathcal{S}}}{\sum}q_{s}u_{s})\}$$
For k=1,....,S
\begin{eqnarray*}
\dfrac{\partial U}{\partial q_{k}} & = & u_{k}\{1-\eta\underset{s\in\mathbf{\mathcal{S}}}{\sum}p_{s}\mu'(u_{s}-\underset{s\in\mathbf{\mathcal{S}}}{\sum}q_{s}u_{s})\}\\
 & = & u_{k}\{1-\eta P_{+}-\eta\lambda(1-P_{+})\}
\end{eqnarray*}
where $P_{+}=\underset{A}{\sum}p_{s},\: A=\{s\in\mathbf{\mathcal{S}}:\; u_{s}-\underset{s\in\mathbf{\mathcal{S}}}{\sum}q_{s}u_{s}\geq0\}$.
\\
For $\dfrac{\partial U}{\partial q_{k}}|_{p_{1,}...p_{S}}\neq0$
(for at least one k), the conclusion holds obviously. 

Consider the case that $\dfrac{\partial U}{\partial q_{k}}|_{p_{1,}...p_{S}}=0$,
$\forall k\in\mathbf{\mathcal{S}}$. \\

$\dfrac{\partial U}{\partial q_{k}}|_{p_{1,}...p_{S}}=0$ holds for all
k, iff $P_{+}=\underset{A^{0}}{\sum}p_{s}=\dfrac{\eta\lambda-1}{\eta(\lambda-1)},\: A^{0}=\{s\in\mathbf{\mathcal{S}}:\; u_{s}-\underset{s\in\mathbf{\mathcal{S}}}{\sum}p_{s}u_{s}\geq0\}$. 

Therefore, we have
\begin{eqnarray*}
U_{BS} & = & \underset{s\in\mathbf{\mathcal{S}}}{\sum}q_{s}u_{s}+\eta\{\underset{s\in A}{\sum}p_{s}u_{s}-(\underset{s\in A}{\sum}p_{s})\cdot\underset{s\in\mathbf{\mathcal{S}}}{\sum}q_{s}u_{s}+\lambda\cdot\underset{s\in A}{\sum}p_{s}u_{s}-\lambda\cdot(\underset{s\in\overline{A}}{\sum}p_{s})\cdot\underset{s\in\mathbf{\mathcal{S}}}{\sum}q_{s}u_{s})\}\\
 & = & \eta\{\underset{s\in A}{\sum}p_{s}u_{s}+\lambda\cdot\underset{s\in\overline{A}}{\sum}p_{s}u_{s}\}
\end{eqnarray*}

The total utility is independent of $q_{s}$ and $U_{BS}\equiv U_{RE}$. $\quad \quad Q.E.D.$
%\end{proof*}
\\ \\
\textbf{Proposition \ref{tradeoff}: (Beliefs Tradeoff among Different States)}\\
Consider the following maximization problem
$$\max _{ q_{ s },s=1,...S }{ \underset { s\in { { S } } }{ \sum   } q_{ s }u_{ s }+\eta \{ \underset { s\in { { S } } }{ \sum   } p_{ s }\mu (u_{ s }-\underset { s\in { { S } } }{ \sum   } q_{ s }u_{ s })\}  }$$
$$s.t. \underset{s\in\mathbf{\mathcal{S}}}{\sum}q_{s}=1.$$\\From $\dfrac{\partial U}{\partial q_{k}}=u_{k}\{1-\eta P_{+}-\eta\lambda(1-P_{+})\}$,
we see that, for $u_{k}>0,$ $U$ is increasing in $q_{k}$ iff $P_{+}>P^{*}$ and is decreasing in $q_{k}$ iff $P_{+}<P^{*}$, where $P^{*}=\dfrac{\eta\lambda-1}{\eta(\lambda-1)}$.\\ 
The constraint $\underset{s\in\mathbf{\mathcal{S}}}{\sum}q_{s}=1$ requires that any increase in $q_{k}$ must come together with decrease(s) in subjective probability(ies) of other states. For simplification, consider the case that $q_{k}$ and $q_{l}\:(k>l)$ change together. Suppose $P_{+}>P^{*}$ and all the other subjective
beliefs are given, since $u_{k}>u_{l}$ for all $k>l$, $\dfrac{\partial U}{\partial q_{k}}>\dfrac{\partial U}{\partial q_{l}}$. A small increase in $q_{k}$ with a small decrease in $q_{l}$ will increase the total utility. Similar analysis for the case $P_{+}<P^{*}$. Our proof replicates but simplifies the proof using Kahn-Tucker condition. $\quad \quad Q.E.D.$
%\end{proof*}
\\ \\
\textbf{Proposition \ref{op}:(Over-optimistic versus Over-pessimistic)}\\
From the proof of Proposition 2, we know that for $P_{+}\neq P^{*}$, further biases are always desirable. Therefore, at optimal
beliefs, $P_{+}=P^{*}=\dfrac{\eta\lambda-1}{\eta(\lambda-1)}$. Since
the optimal $P_{+}$ is uniquely determined by $\eta$ and $\lambda$,
and the objective probabilities are exogenous, the set of $u_{s}$
above the expectation is also uniquely determined. Even
though the value of $\underset{s\in\mathbf{\mathcal{S}}}{\sum}q_{s}u_{s}$
is given, there still exist multiple combinations of $\{q_{s}\}$. As long as sets of $\{q_{s}\}$ generate the required value of $\underset{s\in\mathbf{\mathcal{S}}}{\sum}q_{s}u_{s}$,
they will all achieve the same value of total utility.

Further more, directly from proof of Proposition 2, we see that if
$P_{+}^{0}>(<)P^{*}$, the decision maker will be up-biased(down-biased) in the upper-rank outcomes and down-biased(up-biased) in the lower-rank outcomes.
Therefore, $\underset{s\in\mathbf{\mathcal{S}}}{\sum}q_{s}u_{s}>(<)\underset{s\in\mathbf{\mathcal{S}}}{\sum}p_{s}u_{s}$
iff $P_{+}^{0}>(<)P^{*}$. $\quad \quad Q.E.D.$
%\end{proof*}
\\ \\
\textbf{Proposition \ref{time}: (Information Timing Preference)}\\
Suppose an agent holds optimal subjective beliefs $\{q_{s}\}_{s\in\mathbf{\mathcal{S}}}$. \\
If $i=k$, the agent knows that $Z_{k}$ will happen at t=2. Then, at t=1, $U_{k}^{A}=u_{k}+\eta\mu(u_{k}-\underset{s\in\mathbf{\mathcal{S}}}{\sum}q_{s}u_{s});$ 
at t=2, $U_{k}^{R}=1\times\eta\mu(u_{k}-u_{k})+0\times\underset{s\in\mathbf{\mathcal{S}}/k}{\sum}\eta\mu(u_{s}-\underset{s\in\mathbf{\mathcal{S}}}{\sum}q_{s}u_{s})=0.$ The total utility from observing $i=k$ is, 
$$U_{k}=U_{k}^{A}+U_{k}^{R}=u_{k}+\eta\mu(u_{k}-\underset{s\in\mathbf{\mathcal{S}}}{\sum}q_{s}u_{s}),$$ \\
The agent's prospective utility from getting information in advance is, 
\begin{equation*}
U_{early} =  \underset{s\in\mathbf{\mathcal{S}}}{\sum}q_{s}u_{s}+\eta \{\underset{s\in\mathbf{\mathcal{S}}}{\sum}q_{s}\mu(u_{s}-\underset{s\in\mathbf{\mathcal{S}}}{\sum}q_{s}u_{s})\}.
\end{equation*}\\
Utility without early information is as before, 
\begin{eqnarray*}
U_{wait} & = & \underset{s\in\mathbf{\mathcal{S}}}{\sum}q_{s}u_{s}+\eta\{\underset{s\in\mathbf{\mathcal{S}}}{\sum}p_{s}\mu(u_{s}-\underset{s\in\mathbf{\mathcal{S}}}{\sum}q_{s}u_{s})\}\\
\end{eqnarray*}
Early information is strictly preferred iff $U_{early}>U_{wait}$ holds. Since
\begin{equation*}
U_{early}-U_{wait}  =  \eta\cdot\{\underset{s\in A}{\sum}(q_{s}-p_{s})\cdot(u_{s}-\underset{s\in\mathbf{\mathcal{S}}}{\sum}q_{s}u_{s})+\underset{s\in\overline{A}}{\sum}(q_{s}-p_{s})\cdot\lambda(u_{s}-\underset{s\in\mathbf{\mathcal{S}}}{\sum}q_{s}u_{s})\},
\end{equation*}
$U_{early}>U_{wait}$ iff\\
\begin{equation}\label{timeD}
\underset{s\in A}{\sum}(q_{s}-p_{s})\cdot(u_{s}-\underset{s\in\mathbf{\mathcal{S}}}{\sum}q_{s}u_{s}) > \lambda \underset{s\in\overline{A}}{\sum}(p_{s}-q_{s})\cdot(u_{s}-\underset{s\in\mathbf{\mathcal{S}}}{\sum}q_{s}u_{s}).
\end{equation}\\
For $s\in A,$ where $A=\{s\in\mathbf{\mathcal{S}}:\; u(Z_{s})-\underset{s\in\mathbf{\mathcal{S}}}{\sum}q_{s}u(Z_{s})\geq0\}$,
$(u_{s}-\underset{s\in\mathbf{\mathcal{S}}}{\sum}q_{s}u_{s})\geq0$
by definition; and for $s\in\overline{A},$ $(u_{s}-\underset{s\in\mathbf{\mathcal{S}}}{\sum}q_{s}u_{s})<0$. \\
If $P_{+}^{0}>(<)P^{*}$, by Proposition 3, $q_{s}-p_{s}\geq(\leq)0$
if $s\in A$ while $p_{s}-q_{s}\geq(\leq)0$ if $s\in\overline{A}$,
strict inequality holds for at least one s in each subset. Therefore,
only one of $\underset{s\in A}{\sum}(q_{s}-p_{s})\cdot(u_{s}-\underset{s\in\mathbf{\mathcal{S}}}{\sum}q_{s}u_{s})>0$
and $\underset{s\in\overline{A}}{\sum}(p_{s}-q_{s})\cdot(u_{s}-\underset{s\in\mathbf{\mathcal{S}}}{\sum}q_{s}u_{s})>0$
can hold. \\
When $P_{+}^{0}>P^{*}$, LHS of \ref{timeD} is greater than 0, and $U_{early}>U_{wait}$;
when $P_{+}<P^{*}$, RHS of \ref{timeD} is greater than 0, and $U_{early}<U_{wait}$. $\quad \quad Q.E.D.$
%\end{proof*}
\\ \\
\textbf{Proposition \ref{twoindep},\ref{twoindepgen}: (Two Symmetrically Distributed Lotteries \& Two Lotteries: The General Case)}\\
We prove the general case in Proposition \ref{twoindepgen}. Proposition \ref{twoindep} can be easily derived from Proposition \ref{twoindepgen}.\\
We start with part(ii) in Proposition 6 and prove the case $P^{0}_{+A}>P^*$, $P^{0}_{+B}<P^*$. The case $P^{0}_{+A}<P^*$, $P^{0}_{+B}>P^*$ holds by symmetry. \\
From Proposition \ref{op}, if $P^{0}_{+A}>P^*$ and $P^{0}_{+B}<P^*$, then an agent is on average over-optimistic about the payoff of the lottery $A$, and over-pessimistic about the payoff of the lottery $B$. Therefore, we have, $$
E_{g_{B(\cdot)}^{*}}(Z_{B})<E_{f_{B(\cdot)}}(Z_{B})=E_{f_{A(\cdot)}}(Z_{A})<E_{g_{A(\cdot)}^{*}}(Z_{A}) 
$$
For a naive agent, lottery A is strictly preferred to lottery B.\\
Consider part(i), in which case either both $P^{0}_{+A}$ and $P^{0}_{+B}$ are greater than $P^*$ or both smaller than $P^*$.
Notice that optimal belief $g_i^*(\cdot)$ ensures that $\ensuremath{\underset{a}{\overset{+\infty}{\int}}f_{A}(Z)\mathrm{d}Z=P^{*},}\ensuremath{\underset{b}{\overset{+\infty}{\int}}f_{B}(Z)dZ=P^{*}}$, \\
where $a=min\{Z_{A}:\; Z_{A}-\ensuremath{\underset{-\infty}{\overset{+\infty}{\int}}g_{A}^{*}(Z_{A})Z_{A}\mathrm{d}Z_{A}\geq0}\}=E_{g_{A}^{*}}(Z_{A})$, \\
and $b=min\{Z_{B}:\; Z_{B}-\ensuremath{\underset{-\infty}{\overset{+\infty}{\int}}g_{B}^{*}(Z_{B})Z_{B}\mathrm{d}Z_{B}\geq0}\}=E_{g_{B}^{*}}(Z_{B}) $. \\
Hence, if $a>b$, $E_{g_{A}^{*}}(Z_{A})>E_{g_{B}^{*}}(Z_{B})$ and $\underset{a}{\overset{+\infty}{\int}}[f_{A}(Z)-f_{B}(Z)]\mathrm{d}Z>0. $\\
Similar proof for the case $a<b$. It is easy to derive Proposition \ref{twoindep} from here. $\quad \quad Q.E.D.$
%\end{proof*}
\\ \\
\textbf{Proposition \ref{lotterys}:(Choice between Two Lotteries: Sophisticated Case)}\\
The objective function $U=E_g(Z)+\eta E_f\mu[Z-E_g(Z)]$ can be reformed as, 
\begin{eqnarray*}
U & = & \int_{E_{g}(Z)}^{+\infty}f(Z)Z\mathrm{d}Z+\lambda\int_{-\infty}^{E_{g}(Z)}f(Z)Z\mathrm{d}Z\\
& = & \eta E_{f}(Z)+(1-\eta)E_{g}(Z)+(\lambda-1)\eta\ensuremath{\underset{loss}{\overset{}{\int}}f(Z)}E_{g}(Z)\mathrm{d}Z+(\lambda-1)\eta\ensuremath{\underset{loss}{\overset{}{\int}}f(Z)Z\mathrm{d}Z}.
\end{eqnarray*}
At optimal beliefs, we have $\underset{loss}{\overset{}{\int}}f(Z)\mathrm{d}Z=1-P^*=\dfrac{1-\eta}{\eta(\lambda-1)}$. Substitute $\dfrac{1-\eta}{\eta(\lambda-1)}$ back into the reformed objective function, we have, \\
$$
\eta E_{f}(Z)+(\lambda-1)\eta\ensuremath{\underset{loss}{\overset{}{\int}}f(Z)Z\mathrm{dZ}}
$$
Since $\ensuremath{E_{f_{A}}}(Z_{A})=\ensuremath{E_{f_{B}}}(Z_{B})$, our conclusions hold obviously. $\quad \quad Q.E.D.$
%\end{proof*}
\\ \\
\textbf{Proposition \ref{riskfreenaive}:(Risk Taking due to Optimism and Pessimism: Naive Case)}\\
The agent aims to maximize $\int u(R_{f}+\alpha R)\mathrm{dF(R)}$. The objective function is concave in $\alpha$ as $\int u''(R_{f}+\alpha R)R^{2}\mathrm{dF(R)\leq0}$. For $0\leq \alpha \leq 1$, if $\alpha$ is optimal, it must satisfy the Kuhn-Tucker first order condition: 
$$
\phi(\alpha)=\int u'(R_{f}+\alpha R)R\mathrm{dF(R)\{\begin{array}{c}
\leq0\quad if\;\alpha<1\\
\geq0\quad if\;\alpha>0
\end{array}} . 
$$
Notice that $\int RdF(R)>0$ implies $\phi(0)>0$. Hence, $\alpha=0$ cannot satisfy the first order condition. We conclude that the optimal portfolio has $\alpha>0$. Same analogy can be applied in the case $E(R)<0$. \\
The proof below is presented in the discrete multi-state case to make mathematical expressions clear. All steps are applicable to the continuous distribution. \\
The problem of a biased agent to choose $\alpha^{BS}$  for given $\{q_{s}\}_{s\in\mathcal{S}}$ is \\
$$\underset{\alpha}{Max}\underset{s\in\mathcal{S}}{\sum}q_{s}u(R_{f}+\alpha R_{s}).$$ FOC of this problem is 
\begin{equation*}\label{refoc}
\underset{s\in\mathcal{S}}{\sum}q_{s}u'(R_{f}+\alpha^{BS}R_{s})=0,  
\end{equation*}
where $\alpha^{BS}$ is the optimal allocation of wealth to the risky asset under biased beliefs.\\
We examine the agent's FOC for optimal $\alpha^*$. Consider moving $\mathrm{d}\hat{\omega}$ from state $s'$ to state $s''$ with $R_s''>R_s'$, we have:
\begin{equation*}
(u'(R_f+\alpha^*R_s'')R_s''-u'(R_f+\alpha^*R_s')R_s')\mathrm{d}\hat{\omega}+\sum_{s\in{\mathcal{S}}}\hat{q_s}u''(R_f+\alpha^*R_s)R_s^2\mathrm{d}\alpha^*=0
\end{equation*}
\begin{equation*}
\dfrac{\mathrm{d}\alpha^*}{\mathrm{d}\hat{\omega}}=\dfrac{u'(\cdot)R_s'-u'(\cdot)R_s''}{\sum_{s\in{\mathcal{S}}}q_su''(R_f+\alpha^*R_s)R_s^2}>0. 
\end{equation*}
Therefore, optimal $\alpha^*$ is increasing in the subjective probabilities putting on upper-ranking outcomes. From Proposition \ref{tradeoff}, an optimistic agent is up-biased because she overestimates the probabilities of good outcomes and underestimates the probabilities of bad outcomes. For $\alpha^{BS}>0$, better outcomes refer to the states with higher positive returns, while for $\alpha^{BS}<0$, better outcomes refer to the states with lower negative returns. For $\alpha^{RE}>0$ and $\alpha^{BS}>0$, an optimistic agent is one who overestimates the probabilities on positive returns. To bring down the biased beliefs from optimistic level to rational level, $\alpha$ must decrease. Therefore, we have $\alpha^{OP}>\alpha^{RE}>0$. Instead, a pessimistic agent who overestimates the probabilities on the low returns needs to increase $\alpha$ to get beliefs back to the rational level. Hence, we have $0<\alpha^{PE}<\alpha^{RE}$. 
For $\alpha^{RE}>0$ and $\alpha^{BS}<0$, probabilities on low returns are overestimated by an optimistic agent and $\alpha$ must increase to get beliefs back to the rational level. For this reason, we have $\alpha^{OP}<0<\alpha^{RE}$. For a pessimistic agent with $\alpha^{PE}<0$, $\alpha^{RE}<\alpha^{PE}<0$ leads to a contradiction to the assumption that $\alpha^{RE}>0$. Our proof for $\alpha^{RE}>0$ is completed.
Similar analysis can be applied in the case $\alpha^{RE}<0$. $\quad \quad Q.E.D.$
%\end{proof*}
\\ \\
\textbf{Proposition \ref{riskfreeso}:(Risk Taking due to Optimism and Pessimism: Sophisticated Case)}\\
The problem of choosing optimal $\alpha^{BS}$ for given optimal beliefs $\{q^*_s\}_{s\in \mathcal{S}}$ is, 
$$
\underset{\alpha}{Max}U=\underset{s\in\mathcal{S}}{\sum}q_{s}^{*}u(R_{f}+\alpha R_{s})+\eta\underset{s\in\mathcal{S}}{\sum}p_{s}\mu[u(R_{f}+\alpha R_{s})-\underset{s\in\mathcal{S}}{\sum}q_{s}^{*}u(R_{f}+\alpha R_{s})] 
$$
%At optimal beliefs, we have 
%\begin{equation*}
%U_{BS} = \eta\underset{s\in\mathcal{S}}{\sum}p_{s}u(R_{f}+\alpha R_{s})R_{s}+\eta(\lambda-1)\underset{-BS}{\sum}p_{s}u(R_{f}+\alpha R_{s})R_{s}
%\end{equation*}
We proved in Proposition \ref{riskfreenaive} that if $E(R)>(<)0$ then $\alpha^{RE}>(<)0$.
The first order condition with respect to $\alpha$ at optimal beliefs is
\begin{equation*}
\dfrac{\partial U}{\partial\alpha}\mid_{\alpha, q_{s}=q_{s}^{*}}  = \eta\underset{s\in\mathcal{S}}{\sum}p_{s}u'(\cdot)R_{s}+\eta(\lambda-1)\underset{-BS}{\sum}p_{s}u'(\cdot)R_{s}=0
\end{equation*}
Instead, the first order condition with respect to $\alpha$ for a rational agent is
\begin{equation*}
\dfrac{\partial U}{\partial\alpha}\mid_{\alpha, q_{s}=p_{s}}  = \underset{s\in\mathcal{S}}{\sum}p_{s}u'(\cdot)R_{s}=0.
\end{equation*}
Since $\dfrac{\partial^2 U}{\partial\alpha^2}=\sum { q_{ s }u''(\cdot )R_{ s }^{ 2 } } <0$, if $\alpha^{RE}>0$ and $\eta(\lambda-1)\underset{-BS}{\sum}p_{s}u'(\cdot)R_{s}>0$, then
$$
\dfrac{\partial U}{\partial\alpha}\mid_{\alpha=\alpha^{RE}, q_{s}=q_s^*}>0.
$$
We must have $\alpha^{BS}>\alpha^{RE}>0$. \\
Instead, when $\alpha^{RE}>0$ and $\eta(\lambda-1)\underset{-BS}{\sum}p_{s}u'(\cdot)R_{s}<0$, since 
$$
\dfrac{\partial U}{\partial\alpha}\mid_{\alpha=\alpha^{RE}, q_{s}=q_s^*}<0,
$$
we must have $\alpha^{BS}<\alpha^{RE}$ and the proof for the case $\alpha^{RE}>0$ is completed. Conclusions for $\alpha^{RE}<0$ can be proved by the same logic.

Next, we need to prove the second part of the proposition. First we assume that at optimal beliefs, the subjective expectation is $E_{q_s^*}u(R_f+\alpha R_s)$. Consider moving $\mathrm{d}\hat{\omega}>0$ from state $s'$ to $s''$ with $R_{s''}>R_{s'}$.  Suppose with no change in $\alpha$, the new subjective expectation is $E_{q_s^*}u(R_f+\alpha R_s)+\Delta$, where $\Delta=\mathrm{d}\hat{\omega}(u(R_f+\alpha R_{s''})-u(R_f+\alpha R_{s'}))$. Suppose $\Delta$ is small enough such that only one state $\tilde{s}$ moves from gain to loss due to the increase in expectation. This assumption is always satisfied under the continuous distribution assumption. 
We examine the FOC for the optimal $\alpha$: 
\begin{eqnarray*}
\eta(\lambda-1)p_{\tilde{s}}(\mathrm{d}\hat{\omega})u'(R_{f}+\alpha R_{\tilde{s}})R_{\tilde{s}}+\eta(\lambda-1)\underset{-BS}{\Sigma}p_{s}u''(R_{f}+\alpha R_{s})R_{s}^{2}\}\mathrm{d\alpha}=0 
\end{eqnarray*}

%\eta(\lambda-1)p_{\tilde{s}}(\mathrm{d}\hat{\omega})u'(R_{f}+\alpha R_{\tilde{s}})R_{\tilde{s}}+\{\eta\underset{s\in\mathcal{S}}{\Sigma}p_{s}u''(R_{f}+\alpha R_{s})R_{s}^{2}+\eta(\lambda-1)\underset{-BS}{\Sigma}p_{s}u''(R_{f}+\alpha R_{s})R_{s}^{2}\}\mathrm{d\alpha}=0 
%\end{equation*}

Since $\eta\underset{s\in\mathcal{S}}{\Sigma}p_{s}u''(R_{f}+\alpha R_{s})R_{s}^{2}+\eta(\lambda-1)\underset{-BS}{\Sigma}p_{s}u''(R_{f}+\alpha R_{s})R_{s}^{2}<0$, for $\mathrm{d\alpha}>0$, we must have $\eta(\lambda-1)p_{\tilde{s}}u'(R_{f}+\alpha R_{\tilde{s}})R_{\tilde{s}}>0$ and $R_{\tilde{s}}>0$. 

Instead, if $R_{\tilde{s}}<0$, we need the down-bias $-\mathrm{d}\hat{\omega}$ to move $\eta(\lambda-1)p_{\tilde{s}}u'(R_{f}+\alpha R_{\tilde{s}})R_{\tilde{s}}$ out of the loss region. Notice that the certain state moving into or out of the loss region is the one on the margin of the gain and loss regions. Therefore, 
$$
u(R_f+\alpha R_{\tilde{s}})=E_{q_s^*}u(R_f+\alpha R_s).
$$
With a small change in denotation, we can get $R_{\tilde{s}}=R_{CE}$ in our proposition.

We proved that if $R_{CE}>0$, $\alpha$ is increasing in $\mathrm{d}\hat{\omega}$; if  $R_{CE}<0$, $\alpha$ is decreasing in $\mathrm{d}\hat{\omega}$. The final conclusion and Lemma 3 can be easily derived from here. $\quad \quad Q.E.D.$
%\end{proof*}

%When $R_{\tilde{s}}>0$, to bring down the beliefs back to rational level, an optimistic investor will adjust $\alpha$ to a lower level, therefore, $\alpha^{RE}<\alpha^{OP}$. 

%Instead, from pessimistic to rational, the investor need to move up her subjective beliefs. If $R_{\tilde{s}}>0$, then she has $\alpha^{RE}>\alpha^{PE}$. 

\section{Mathematical Appendix B}
Appendix B presents the model in Chapter 3 by relaxing the linear restriction on the gain-loss utility $\mu(x)$. \\
Specifically, we assume
$$
\mu '( - x) = \lambda (x)\mu '(x),
$$\\
where $\lambda (x) > 1,x > 0$, and $\mathop {\lim }\limits_{x \to 0} \lambda (x) = 1$, $\mu ''(x) \le 0$,  $\lambda '(x)\leq0$.
Conclusions under the new assumption are similar to those in Chapter 3. \\
All the optimal conditions below are derived from the FOC: 
${{\partial U} \over {\partial q_s }} = 0$, where
$$
U = \mathop \Sigma \limits_{s \in {\cal S}} q_s^{} u(Z_s^{} ) + \eta \mathop \Sigma \limits_{s \in {\cal S}} p_s^{} \mu [u(Z_s^{} ) - \mathop \Sigma \limits_{s \in {\cal S}} q_s^{} u(Z_s^{} )].
$$
Previous conclusions under linear assumption in Proposition 3 are:

For $\mu (x) = x\quad (x > 0),\mu (x) = \lambda x\quad (x < 0)$,
\begin{itemize}
\item[(i)]{Linear: Two-state Case} \\
$P^*  = {{\eta \lambda  - 1} \over {\eta (\lambda  - 1)}} = 1 + {{\eta  - 1} \over {\eta (\lambda  - 1)}}$ is independent of $q$. \\
Optimal q is: $q^*=0$, if $p<P^*$; $q^*=1$, if $p>P^*$. 
\item[(ii)] {Linear: S-state Case}\\
Optimal $\{q^*_s\} _{s \in {\cal S}}$ satisfy: $
P_ +   = P^*  = {{\eta \lambda  - 1} \over {\eta (\lambda  - 1)}},$ where $P_ +   = \mathop \Sigma \limits_A p_s ,A = \{ s \in {\cal S}:\>u(Z_s ) - \mathop \Sigma \limits_{s \in {\cal S}} q_s u(Z_s ) \ge 0\} 
$, and $P^*$ is independent of $q_s$. Therefore, there exist more than one optimal set of $
\{ q^*_s\} _{s \in {\cal S}} 
$, and they all achieve a certain value of $\mathop \Sigma \limits_{s \in {\cal S}} q^*_s u(Z_s )$(for given$\{ u(Z_s )\} _s^{}$), which is uniquely determined by $P^*  = {{\eta \lambda  - 1} \over {\eta (\lambda  - 1)}}$ for given $\{ p_s^{} \} _{s \in {\cal S}}$. (Conclusions from Proposition 3). 
\end{itemize}

Next, we discuss the model under new assumptions.\\
We assume $\mu '( - x) = \lambda (x)\mu '(x)$, where $\lambda (x) > 1,x > 0$, and $\mathop {\lim }\limits_{x \to 0} \lambda (x) = 1$, $\mu ''(x) \le 0$, $\lambda'(x)\leq 0$.
\begin{itemize}
\item[(i)]{General: Two-state Case}\\ 
Suppose we have two states with $u(1)=1, u(0)=0$.\\
From FOC, we can derive $
p = 1 + {{\eta \mu '(1 - q) - 1} \over {\eta [\mu '(q)\lambda (q) - \mu '(1 - q)]}}$\\
Since $0<p<1$ and ${{\eta \mu '(1 - q) - 1} \over {\eta [\mu '(q)\lambda (q) - \mu '(1 - q)]}} < 0$,  the denominator and the numerator must have opposite signs. \\
For $\lambda '(x) \le 0
$, the cut-off value ${{\eta \mu '(1 - q) - 1} \over {\eta [\mu '(q)\lambda (q) - \mu '(1 - q)]}}$ is negative and decreasing in q. Therefore, for $p > 1 + {{\eta \mu '(1 - p) - 1} \over {\eta [\mu '(p)\lambda (p) - \mu '(1 - p)]}}$, the optimal $q^*=1$, otherwise, optimal $q^*=0$. \\
To be more specific, let $\mu '(x) = \beta 
$, then \\
$$
p = {{\eta \beta \lambda (q) - 1} \over {\eta \beta [\lambda (q) - 1]}} = 1 + {{\eta \beta  - 1} \over {\eta \beta [\lambda (q) - 1]}}.
$$
For $0<p<1$, we must have $
\eta \beta  < 1
$. \\
If $
\lambda '(q) \le 0
$, then $
{{\eta \beta \lambda (q) - 1} \over {\eta \beta [\lambda (q) - 1]}}
$ is decreasing in q. Optimal $q^*=1$ if $
p > {{\eta \beta \lambda (p) - 1} \over {\eta \beta [\lambda (p) - 1]}}
$ and $q^*=0$ otherwise. \\
\item[(ii)]{General: S-state Case}\\
From FOC, we have $$
\eta \{ \sum\limits_{Gain} {p_s^{} } \mu '(u_s^{}  - \sum\limits_S {q_s^{} u_s^{} } ) + \sum\limits_{Loss} {p_s^{} } \mu '(\sum\limits_S {q_s^{} u_s^{} }  - u_s^{} )\lambda (\sum\limits_S {q_s^{} u_s^{} }  - u_s^{} )\}  = 1.
$$
Specifically, for $
\mu '( \cdot ) \equiv \beta 
$, $
\sum\limits_{Loss} {p_s^{} } [\lambda (\sum\limits_S {q_s^{} u_s^{} }  - u_s^{} ) - 1] = {{1 - \eta \beta } \over {\eta \beta }}
.$ \\
Since the $LHS>0$, we have $\eta \beta<1$. \\
For $\lambda '(\cdot)\leq0$, if $
\sum\limits_{Loss} {p_s^{} } [\lambda (\sum\limits_S {q_s^{} u_s^{} }  - u_s^{} ) - 1] < {{1 - \eta \beta } \over {\eta \beta }}
$, then the total utility is increasing in $
\sum\limits_S {q_s^{} u_s^{} } 
$. An increase in $
\sum\limits_S {q_s^{} u_s^{} } 
$ will decrease the value of $
\sum\limits_{Loss} {p_s^{} } [\lambda (\sum\limits_S {q_s^{} u_s^{} }  - u_s^{} ) - 1]
$(if the range of loss remains the same), making it further below $
{{1 - \eta \beta } \over {\eta \beta }}
$. Therefore, optimal sets of $\{q^*_s\}$ are those satisfy $$
\sum\limits_{Loss} {p_s^{} } [\lambda (\sum\limits_S {q^*_s u_s}  - u_s) - 1] = {{1 - \eta \beta } \over {\eta \beta }},
$$
where $
Loss = \{ s \in {\cal S}:\>u_s-\sum\limits_{s \in {\cal S}}{q^*_s u_s}< 0\} 
$. 
\end{itemize}

\bibliographystyle{aea}
\bibliography{ref0415}

\end{document}